\documentclass[epsf,prb,twocolumn,showpacs,nofootinbib]{revtex4-1}

\usepackage[pdftex]{graphicx}
\usepackage{dcolumn}
\usepackage{bm}
\usepackage{epsfig}
\usepackage{latexsym}
\usepackage{amsmath}
\usepackage{amsfonts}
\usepackage{amssymb}
\usepackage{color}
\usepackage{array}
\usepackage{framed}

\begin{document}

\title{Generalized Electromagnetism of Subdimensional Particles: \\A Spin Liquid Story}
\author{Michael Pretko\\
\emph{Department of Physics, Massachusetts Institute of Technology, Cambridge, MA 02139, USA}}
\date{June 21, 2017}

\begin{abstract}
It has recently been shown that there exists a class of stable gapless spin liquids in 3+1 dimensions described by higher rank tensor $U(1)$ gauge fields, giving rise to an emergent tensor electromagnetism.  The tensor gauge field of these theories couples naturally to subdimensional particles (such as fractons), which are restricted by gauge invariance to move only along lower-dimensional subspaces of the system.  We here work out some of the basic generalized electromagnetic properties of subdimensional particles coupled to tensor electromagnetism, such as generalized electrostatic fields, potential formulations, Lorentz forces, Maxwell equations, and Biot-Savart laws.  Some concepts from conventional electromagnetism will carry over directly, while others require significant modification.
\end{abstract}

\maketitle

\section{Introduction}

Spin liquids are fascinating states of matter which exhibit long-range entanglement in the ground state and an exotic spectrum of emergent excitations.\cite{lucile}  As a simple example, a conventional three-dimensional $U(1)$ quantum spin liquid is described by the familiar $U(1)$ gauge theory of Maxwell electromagnetism.  Such a system has an emergent gapless photon (with the effective ``speed of light" determined by microscopic parameters), as well as emergent charged particles, coming in both electric and magnetic varieties.  The gaplessness of the emergent photon is protected by gauge invariance, which makes these phases stable against perturbations to the Hamiltonian, without the need for symmetry protection.

This gauge structure of the theory gives rise to an emergent electromagnetism in $U(1)$ spin liquids.  This quantum electromagnetic theory can be conveniently described in terms of a vector potential $\vec{A}$, where the magnetic field is given by $\vec{B} = \nabla\times\vec{A}$.  The electric field $\vec{E}$ plays the role of the canonical conjugate to $\vec{A}$, since $\vec{E} = \partial \mathcal{L}_{Max}/\partial\dot{\vec{A}}$.  Microscopically, this gauge structure can be obtained by rewriting the fundamental degrees of freedom, such as spins on a lattice, into a language which maps a geometrically frustrating spin interaction (e.g. ``spin ice rules") into a gauge constraint.  The reader is referred to Reference \onlinecite{u1} for a more thorough discussion of the $U(1)$ quantum spin liquid and its Hilbert space.

While the vector potential theory is most familiar, it has recently been realized that there is a wide class of other spin liquids described by emergent tensor gauge fields.  In particular, it has been shown that systems described by symmetric tensor $U(1)$ gauge fields exhibit a stable deconfined phase, making these a new class of stable gapless spin liquids.\cite{alex}  As in the conventional $U(1)$ spin liquid, these systems have robust gaplessness without needing to rely on symmetry protection.  Furthermore, it has been shown that the particles carrying the gauge charge in these theories obey extra conservation laws which restrict their motion to lie on certain lower-dimensional subspaces.\cite{me}  For example, certain theories have particles carrying a vector charge, which can move only along one-dimensional subspaces, in the direction of their charge vector.  In other models, the fundamental charges are restricted to a zero-dimensional subspace, $i.e.$ they cannot move at all without the creation of additional particles.  Other types of subdimensional particles are also possible.  This subdimensional behavior makes these theories the natural $U(1)$ analogue of the discrete ``fracton" models constructed by Vijay, Haah, and Fu.\cite{fracton1,fracton2}  The same sort of restricted mobility was also seen in earlier work by Chamon and others.\cite{chamon,bravyi,cast,yoshida,haah,haah2}  Such fracton phases have seen a flurry of recent activity.\cite{williamson,han,sagar,prem,hsieh,slagle}

Phases described by symmetric tensor $U(1)$ gauge fields also exhibit an emergent electromagnetism, but of a more exotic form.  The electric and magnetic fields are now tensor-valued objects instead of vectors.  We will see that even currents in these theories must be promoted to tensor objects.  Furthermore, in each distinct higher rank $U(1)$ spin liquid phase, there is a different set of generalized Maxwell equations.  This gives us many new types of generalized electromagnetism to play with.  As discussed in previous work, there are actually an infinite number of such higher rank $U(1)$ theories, by considering tensors of arbitrary rank.  For the sake of finiteness, we shall here work with the rank 2 symmetric tensor theories, of which there are four known distinct spin liquid phases.  Any other higher rank theory ought to be amenable to the same sort of analysis.  We will focus on generalizing some of the basic concepts of undergraduate electromagnetism to these higher rank $U(1)$ theories, such as electrostatic fields, potential formulations, Lorentz forces, and Biot-Savart laws.  Our analysis will mainly focus on the macroscopic picture, abstracting from any specific microscopic models.  For completeness, however, we will review the concrete lattice models in Appendix D.

Before proceeding to the main analysis, we first quickly recap the properties of the four types of rank 2 symmetric tensor $U(1)$ spin liquids, developed in References \onlinecite{alex} and \onlinecite{me}.  In all cases, the degrees of freedom are those of a symmetric rank 2 compact $U(1)$ tensor $A_{ij}$, with a conjugate variable $E_{ij}$ representing a generalized electric field.  As discussed in these references, each phase can be fully specified by the structure of its Gauss's law constraint, and the four types of rank 2 theories are distinguished by the different choices of Gauss's laws available.  The different theories are as follows:

\subsubsection*{A:   Vector Charge Theory}

In this theory, the Gauss's law takes the form $\partial_iE^{ij} = \rho^j$, for vector-valued charge $\rho^j$.  The charges of this theory obey two constraints:
\begin{equation}
\int\vec{\rho}\, = \textrm{constant}\,\,\,\,\,\,\,\,\,\,\,\,\,\,\int\vec{x}\times\vec{\rho}\, = \textrm{constant}
\label{cons}
\end{equation}
reflecting the conservation of both the charge and its angular moment.  In order to obey these conservation laws, the vector charges are forced to become 1-dimensional particles, hopping only along the direction of their charge vector.  The gapless gauge mode for this theory has three independent polarizations.  This theory has a magnetic field tensor $B_{ij}$ which is also symmetric and obeys a corresponding magnetic Gauss's law, $\partial_iB^{ij} = \tilde{\rho}^j$, for magnetic charge $\tilde{\rho}^j$.  The magnetic field tensor has two spatial derivatives, leading to a quadratic dispersion for the gauge mode.  This theory possesses a self-duality between the electric and magnetic tensors.  We shall therefore focus mainly on the case where only electric charges and currents are present in the theory.  Results for magnetic charges and currents can be obtained straightforwardly by duality.

\subsubsection*{B:   Traceless Vector Charge Theory}

We here have the same Gauss's law from the previous theory, $\partial_iE^{ij} = \rho^j$, but also add in a tracelessness constraint, $E^i_{\,\,\,i} = 0$.  We are free to impose this constraint exactly on the entire Hilbert space.  Alternatively, one could consider allowing the constraint to break by adding trace charges, $E^i_{\,\,\,i} = \rho_{tr}$.  This could be treated by similar methods, but would make the analysis more cumbersome and does not change the results qualitatively, so we focus on the case where tracelessness is exact.  This theory has both of the conservation laws of the previous theory, plus two new conservation laws related to the tracelessness:
\begin{equation}
\int(\vec{\rho}\cdot \vec{x})\, = \textrm{constant}\,\,\,\,\,\,\,\,\,\,\,\,\,\,\,\,\,\int[(\vec{x}\cdot\vec{\rho}) \vec{x} - \frac{1}{2}x^2\vec{\rho}] \,= \textrm{constant}
\end{equation}
These conservation laws restrict the fundamental charges from moving at all, turning them into fractons (0-dimensional particles).  The only mobile particles in this theory are bound states, which will be discussed later.  The gapless gauge mode of this theory has two independent polarizations.  The theory has a symmetric traceless magnetic field tensor $B_{ij}$, now with three spatial derivatives, leading to a cubic dispersion for the gauge mode.  Once again, the theory exhibits self-duality, swapping the role of the electric and magnetic field tensors.

\subsubsection*{C:   Traceless Scalar Charge Theory}

We now consider a theory with a two-derivative Gauss's law, $\partial_i\partial_jE^{ij} = \rho$, for scalar charge $\rho$.  Let's also first suppose the electric field tensor is traceless, $E^i_{\,\,\,i} = 0$, saving the traceful analogue for last.  The theory has three constraints on the charge:
\begin{equation}
\begin{split}
\int \rho = \textrm{constant}\,\,\,\,\,\,\,\,\,\,\,\,\int \vec{x}\rho =& \textrm{constant} \\
\int x^2\rho = \textrm{constant}&
\end{split}
\end{equation}
reflecting the conservation of charge, dipole moment, and one specific component of the quadrupole moment.  The fundamental charges are fractons, unable to hop in any direction without the creation of additional charges.  The mobile excitations are dipolar bound states, which behave as two-dimensional particles, only able to hop transversely to the dipole moment.  The gapless gauge mode has four independent polarizations.  The magnetic field tensor $B_{ij}$ is a symmetric traceless tensor, just like $E_{ij}$, leading once again to self-duality.  There is only one spatial derivative in $B_{ij}$, leading to a linear dispersion for the gauge mode.

\subsubsection*{D:   Scalar Charge Theory}

Lastly, we take the Gauss's law to be the same as the previous theory, $\partial_i\partial_jE^{ij} = \rho$, but without imposing any tracelessness constraint.  In some sense, this is actually the simplest of the theories.  However, a few extra comments are necessary, since this theory does not have the self-duality property of the previous theories.  (Thanks are due to Sagar Vijay, who first noticed the issue with duality in this model.)  The electric charges of the theory have two constraints:
\begin{equation}
\int \rho = \textrm{constant}\,\,\,\,\,\,\,\,\,\,\,\,\,\,\,\,\,\,\,\,\int \vec{x}\rho = \textrm{constant}
\end{equation}
corresponding to the conservation of charge and dipole moment.  Once again, the fundamental charges are fractons.  The dipolar bound states of this theory are fully mobile, possessing both longitudinal and transverse motion.  The gapless gauge mode has five independent polarizations.

\begin{table*}[t]
  \begin{tabular}{ | c | c | c | c | c |}
    \hline
    Gauss's Law(s) & Magnetic Tensor, $B_{ij}$ &  Gauge Dispersion & Polarizations & Self-Dual? \\ \hline
    $\partial_i\partial_j E^{ij} = \rho$ & $\epsilon_{iab}\partial^a A^b_{\,\,\,j}$ & $\omega \propto k$ & 5 & No \\ \hline
    $\partial_i\partial_j E^{ij} = \rho,\,\,\,\,E^i_{\,\,\,i}=0$ & $\frac{1}{2}(\epsilon_{iab}\partial^a A^b_{\,\,\,j} + \epsilon_{jab}\partial^a A^b_{\,\,\,i})$ & $\omega\propto k$ & 4 & Yes \\ \hline
    $\partial_iE^{ij} = \rho^j$ & $\epsilon_{iab}\epsilon_{jcd}\partial^a\partial^c A^{bd}$ & $\omega\propto k^2$ & 3 & Yes \\ \hline
    $\partial_iE^{ij} = \rho^j,\,\,\,\,E^i_{\,\,\,i}=0$ & (See Equation \ref{monster}) & $\omega\propto k^3$ & 2 & Yes \\
    \hline
  \end{tabular}
  \caption{Summary of the rank 2 $U(1)$ spin liquids.}
  \label{table}
\end{table*}

Unlike the previous theories, the appropriate magnetic field tensor for this theory is actually a \emph{non-symmetric} (and traceless) tensor $B_{ij}$, with one spatial derivative, leading to linear gauge mode dispersion.  It can readily be checked that the non-symmetric traceless tensor $B_{ij} = \epsilon_{iab}\partial^aA^b_{\,\,\,j}$ represents the stable fixed point of the theory where all five polarizations have the same dispersion.  This non-symmetric tensor obeys a different Gauss's law, $\partial_iB^{ij} = \tilde{\rho}^j$, with vector magnetic charge.  The magnetic charges obey two constraints:
\begin{equation}
\int\vec{\tilde{\rho}} = \textrm{constant}\,\,\,\,\,\,\,\,\,\,\,\,\,\,\,\,\,\int\vec{\tilde{\rho}}\cdot\vec{x} = \textrm{constant}
\end{equation}
which makes the magnetic vector charges 2-dimensional particles, only hopping transversely to their charge vector.

This theory lacks any sense of self-duality between the electric and magnetic sectors.  Nevertheless, it remains stable against confinement.  It is interesting to note that there is a duality between a traceful symmetric tensor description (with one particle and five gauge mode degrees of freedom) and a traceless non-symmetric tensor description (with three particle and five gauge mode degrees of freedom).

Some basic properties of the four phases are summarized in Table \ref{table}.  In the following sections, we will go through each of the four theories one by one.  We will start with the scalar charge theory, since it turns out to be the simplest.  The analysis of this theory in Section 2 will lay the groundwork for discussing the other rank 2 theories.  Most of the important concepts will be developed in this section and will be extended in natural fashion in the following sections.  The casual reader may therefore wish to focus primarily on section 2.

We note that one might also consider theories with ``curl" constraints, such as $\epsilon^{ijk}\partial_jE_k^{\,\,\,\ell} = \rho^{i\ell}$.  These theories are slightly different, as they do not host point particles, and they may have issues with stability.  We therefore relegate a discussion of such constraints to Appendix C.

\section{Scalar Charge Theory}

\subsection{Electrostatic Fields}

The generalized Gauss's law of this theory is given by:
\begin{equation}
\partial_i\partial_j E^{ij} = \rho
\end{equation}
for scalar charge $\rho$.  The fundamental charges in this theory are fractons, unable to hop in any direction without the creation of additional particles.  When such a charge is isolated (a situation which is possible to create), it will provide a delta function source for Gauss's law:
\begin{equation}
\partial_i \partial_j E^{ij} = q\delta^{(3)}(r)
\end{equation}
where charges are quantized as multiples of $q$.  We now wish to know the expectation value of the electric field due to this point source, $\langle E^{ij}\rangle$.  To avoid clutter, we will omit brackets throughout, simply writing $E^{ij}$.  Equivalently, the following analysis can be taken to be applied to the classical limit of the theory.  Since, in the Coulomb phase, the particles can be regarded as independent excitations, and since the low-energy effective theory for the gauge field is rotationally invariant, the generalized electric field may only depend on rotationally invariant quantities.  Furthermore, by dimensional analysis, we know that $E^{ij}$ must scale as $q/r$.  The only such symmetric rank 2 quantities are as follows:
\begin{equation}
E^{ij} = q\bigg(\alpha \frac{\delta^{ij}}{r} + \beta \frac{r^ir^j}{r^3}\bigg)
\label{generalcoul}
\end{equation}
As a first condition, we must satisfy the Gauss's law.  We have:
\begin{equation}
\begin{split}
&\partial_i E^{ij} = q(\beta-\alpha) \frac{r^j}{r^3} \\
\partial_i\partial_j& E^{ij} = 4\pi q(\beta-\alpha)\delta^{(3)}(r)
\end{split}
\end{equation}
so we require $\beta -\alpha = 1/4\pi$ to satisfy Gauss's law.  The electric field then becomes:
\begin{equation}
E^{ij} = q\bigg(\alpha\frac{\delta^{ij}}{r} + \bigg(\alpha + \frac{1}{4\pi}\bigg) \frac{r^ir^j}{r^3}\bigg)
\label{coulomb}
\end{equation}
Note that, unlike the case of conventional electromagnetism, the Coulomb field of a static point charge has not been uniquely specified by Gauss's law and rotational symmetry.  In order to further constrain the electric field, we must resort to another of the generalized Maxwell equations.  For the traceful scalar charge theory, the correct magnetic field tensor is the (non-symmetric) tensor:
\begin{equation}
B^{ij} = \epsilon^{iab}\partial_a A_b^{\,\,\,j}
\end{equation}
The equation governing the evolution of the magnetic field is then:
\begin{equation}
\partial_t B^{ij} = \epsilon^{iab}\partial_a \partial_t A_b^{\,\,\,j} = \epsilon^{iab}\partial_a E_b^{\,\,\,j}
\end{equation}
where we have used the fact that $E_{ij}$ is the canonical momentum to $A_{ij}$ to derive a generalized Faraday's equation.  For a magnetostatic solution, we will then require our Coulomb field to satisfy:
\begin{equation}
\epsilon^{iab}\partial_a E_b^{\,\,\,j} = q(\alpha+\beta)\frac{\epsilon^{ija}r_a}{r^3} = 0
\end{equation}
which means we need $\beta = -\alpha$.  When combined with our earlier condition, $\beta = \alpha+1/4\pi$, we obtain $\alpha = -\frac{1}{8\pi}$ and $\beta = \frac{1}{8\pi}$, so the final result for the static Coulomb field of a point charge (electric monopole) of strength $q$ is:
\begin{equation}
E_{mono}^{ij} = \frac{q}{8\pi}\bigg(\frac{r^ir^j}{r^3} - \frac{\delta^{ij}}{r}\bigg)
\label{monopole}
\end{equation}
Since the differential equations involved have been linear, we can then find the electric field of a general charge distribution by taking superpositions.  In particular, for a dipole of strength and direction $p^i$, the appropriate electric field is $- p^k\partial_k E_{q=1}^{ij}$, which is given by:
\begin{equation}
E^{ij}_{dip} = -\frac{1}{8\pi}\bigg(\frac{\delta^{ij}(p\cdot r)}{r^3} + \frac{(p^ir^j + r^ip^j)}{r^3} - 3\frac{r^ir^j(p\cdot r)}{r^5}\bigg)
\label{dipfield}
\end{equation}

\subsection{Potential Formulation}

There is actually a conceptually cleaner and simpler way to derive these electric field solutions.  In the spirit of normal electromagnetism, we will seek a potential formulation, to mitigate the proliferation of indices.  From our magnetostatic constraint, $\epsilon^{iab}\partial_a E_b^{\,\,\,j} = 0$, we can immediately conclude that $E_{ij} = \partial_i \lambda_j$ for some vector $\lambda_j$.  However, by the symmetry of $E_{ij}$, we must have $\partial_i\lambda_j = \partial_j\lambda_i$.  This implies that the curl of $\lambda$ vanishes, $\epsilon^{ijk}\partial_j \lambda_k = 0$, so $\lambda$ will in turn be a derivative, $\lambda_j = \partial_j \phi$, for some scalar potential $\phi$.  We then have:
\begin{equation}
E_{ij} = \partial_i\partial_j \phi
\end{equation}
(Note that we have not introduced a negative sign, as one would have done in conventional electromagnetism.  The naturalness of this sign convention will be seen shortly.)  This scalar potential significantly reduces the complexity of the problem.  In order to satisfy Gauss's law for a point charge, we must have:
\begin{equation}
\partial_i\partial_j E^{ij} = (\partial^2)^2\phi = q\delta^{(3)}(r)
\end{equation}
By dimensional analysis, $\phi$ must scale as $r$.  The only possibility is $\phi = Cr$ for constant $C$ (it can readily be checked that possible logarithmic terms cannot solve the Gauss's law and can be ruled out).  Differentiating yields $(\partial^2)^2 (Cr) = -8\pi C \delta^{(3)}(r)$, so we require $C = -q/8\pi$.  Then, taking the appropriate derivatives, we immediately obtain Equation \ref{monopole} for the Coulomb field, as expected.  Again, due to the linearity of the differential equations, we can then apply the superposition principle to the potential, instead of to the field directly, which is much simpler.  For example, the potential due to a dipole $p^i$ is given by $\phi_{dip} = -p^i \partial_i C_{q=1}r$.  The scalar potentials for the electric monopoles and dipoles are:
\begin{equation}
\phi_{mono} = -\frac{qr}{8\pi}\,\,\,\,\,\,\,\,\,\,\,\,\,\,\,\phi_{dip} = \frac{(p\cdot r)}{8\pi r}
\label{potent}
\end{equation}
This potential is not simply a convenient mathematical tool.  Just like in conventional electromagnetism, the potential plays a direct physical role.  Consider the energy stored in the electric field of a static charge configuration:
\begin{equation}
\begin{split}
\epsilon = \frac{1}{2}\int E^{ij}E_{ij} = \frac{1}{2}\int E^{ij} \partial_i\partial_j \phi = \\
\frac{1}{2} \int \partial_i\partial_j E^{ij}\phi = \frac{1}{2}\int \rho \phi
\end{split}
\end{equation}
This equation is exactly the same as is obtained in normal electrostatics.  (Note the importance of the sign convention for the potential.)  This tells us that the potential represents the energy associated with a particle at a particular location.  Even though fractons do not possess a conventional sense of forces or equations of motion, we see that fractons nevertheless have a potential energy.  Note that the factor of $1/2$ prevents overcounting between charge pairs.  Interestingly, $\phi$ for a point charge vanishes at the charge's location, so there is no ``self-energy" contribution to the energy integral.  All energy can be viewed as arising from each particle interacting with the potential of the other particles, but not its own.  Since $\phi$ for a point charge grows linearly, we see that separating a group of fractons requires an energy linear in the separation, as has been found in previous work.  This large energy cost naively would suggest that the particles are confined.  However, the immobile nature of fractons, coupled with the fact that the energy density of the electric field is bounded, stabilizes them against collapsing back into the vacuum once the energy cost has been paid to create them.  Thus, the fractons are in fact well-defined excitations, albeit very energetically costly, in a situation reminiscent of vortices in a superfluid.  (See Reference \onlinecite{me} for a more detailed discussion of this ``electrostatic confinement" issue.)

\subsection{Lorentz Force}

While the results of the previous sections are good to have in hand, the knowledge of these electric fields and potentials will not mean much unless we know how the charges of the theory will respond to them.  For isolated fundamental charges of the theory, we already know the answer to this question: they don't respond at all.  The isolated electric monopoles in this theory are fractons, so they cannot hop without a huge input of energy to create extra particles.  They therefore have no equations of motion and do not respond at all to the electromagnetic fields.  However, the dipolar bound state of a positive and negative charge will be freely propagating in this theory, since dipole motion will preserve the global dipole moment, as long as the dipole does not change orientation.  A dipole can therefore respond to the electromagnetic fields, but it cannot change its orientation, except through interaction with other particles.  We can therefore effectively treat an isolated dipole like a freely hopping point particle.

But what is the effect of the fields on this effective particle?  We can draw our intuition from the lattice models for the higher rank spin liquids\cite{alex,cenke1,cenke2}, discussed in Appendix D.  In these models, $A_{ij}$ represents the phase picked up by hopping a $-i$ oriented dipole in the $j$ direction, and also the phase for hopping a $-j$ oriented dipole in the $i$ direction.  Effectively, a dipole $p^j$ responds to the magnetic field tensor just like a conventional charged particle would respond to an ordinary electromagnetic field, with an effective vector potential given by $A_{eff}^j = -\hat{p}_i A^{ij}$ and effective magnetic field $B_{eff}^i = -\epsilon^{ijk}\partial_j \hat{p}^\ell A_{\ell k} = -B^{ij}\hat{p}_j$.  (As a reminder, while $A$ and $E$ are symmetric tensors, $B$ is \emph{not} symmetric, so the index of contraction is quite important here.)  The corresponding effective electric field is $E_{eff}^j = -\hat{p}_i E^{ij}$.  The generalized Lorentz force on a dipole $p^i$ moving with velocity $v^i$ is then given by:
\begin{equation}
F^j = -p_i (E^{ij} + \epsilon^{j\ell k}v_\ell B_k^{\,\,\,i})
\end{equation}
Let us now suppose that a dipole has been placed in an electric field created by some static electric charge distribution, so that $B_{ij} = 0$.  The corresponding electric force is given by:
\begin{equation}
F^j = -p_i\partial^i\partial^j \phi
\end{equation}
where $\phi$ is the scalar potential.  Let us now calculate the work necessary to move the dipole from point 1 to point 2 against the field:
\begin{equation}
W = -\int_1^2 F^j dx_j = \int_1^2 dx^j \partial_j(p^i\partial_i \phi) = (p^i\partial_i\phi)_2 - (p^i\partial_i\phi)_1
\end{equation}
We therefore see that the potential energy associated with dipole $p^i$ is given by $V = p^i\partial_i \phi$, which we could have predicted based on our previous discussion of the potential, but it is nice to see this conclusion arise independently.

Note that the Lorentz force which we have found in this section has a negative sign in front, which looks peculiar at first.  As a sanity check, let us calculate the electric force between two identical dipoles $p^i$.  Making use of the electric field found in Equation \ref{dipfield}, we have that the electric force on a dipole at location $r^i$ due to an identical dipole at the origin is given by:
\begin{equation}
F^j = \frac{1}{8\pi}\bigg(\frac{2p^j(p\cdot r)}{r^3} + \frac{p^2r^j}{r^3} - 3\frac{(p\cdot r)^2r^j}{r^5}\bigg)
\end{equation}
The radial component of this force is:
\begin{equation}
\begin{split}
F^j\hat{r}_j = \frac{1}{8\pi}\bigg(-&\frac{2(p\cdot r)^2}{r^4} + \frac{p^2}{r^2}\bigg) = \\
\frac{p^2}{8\pi r^2}& (1 - \cos^2\theta) = \\
&\frac{p^2\sin^2\theta}{8\pi r^2}
\end{split}
\end{equation}
where $\theta$ is the angle between $p^i$ and $r^i$.  Note that the radial force is always non-negative, indicating a repulsive force between like charges.  Flipping the direction of one of the dipoles would result in an overall negative sign, so two oppositely oriented dipoles will always attract, a state of affairs which makes intuitive sense, since they can recombine into the vacuum.  Interestingly, the Lorentz force between two dipoles vanishes when they line up along a line, such that $(p\cdot r) = pr$.  This corresponds to a minimum of the potential for like dipoles and a maximum of the potential for opposite dipoles.  Therefore, like dipoles energetically prefer to arrange themselves end to end, whereas opposite dipoles prefer to be side by side.

One last comment is in order regarding our Lorentz force on dipoles.  The force increases linearly with the dipole moment $p^i$.  This seems to indicate that there is a larger force on two charges separated by a large distance than two charges right next to each other.  This seems puzzling, since we expect that two well-separated charges should approach the behavior of isolated fractons, which should not move at all.  The resolution comes from the fact that we have identified the force based on the phases associated with hopping matrix elements, but we have not yet accounted for the \emph{magnitude} of the hopping elements.  In other words, we have not accounted for the effective mass of the dipoles.  The magnitude of hopping matrix elements will be much smaller (and the effective mass correspondingly much larger) for dipoles of large separation.  The effective mass of a dipole grows exponentially in the particle separation, $m(p) \propto e^{\alpha p}$, where $\alpha$ is determined by microscopic parameters (via perturbation theory).  Thus, while well-separated dipoles experience an algebraically large force, they have an exponentially large effective mass, resulting in exponential suppression of typical velocities.  In this manner, well-separated dipoles will smoothly approach the limit of fractonic behavior.

\begin{figure*}[t!]
 \centering
 \includegraphics[scale=0.25]{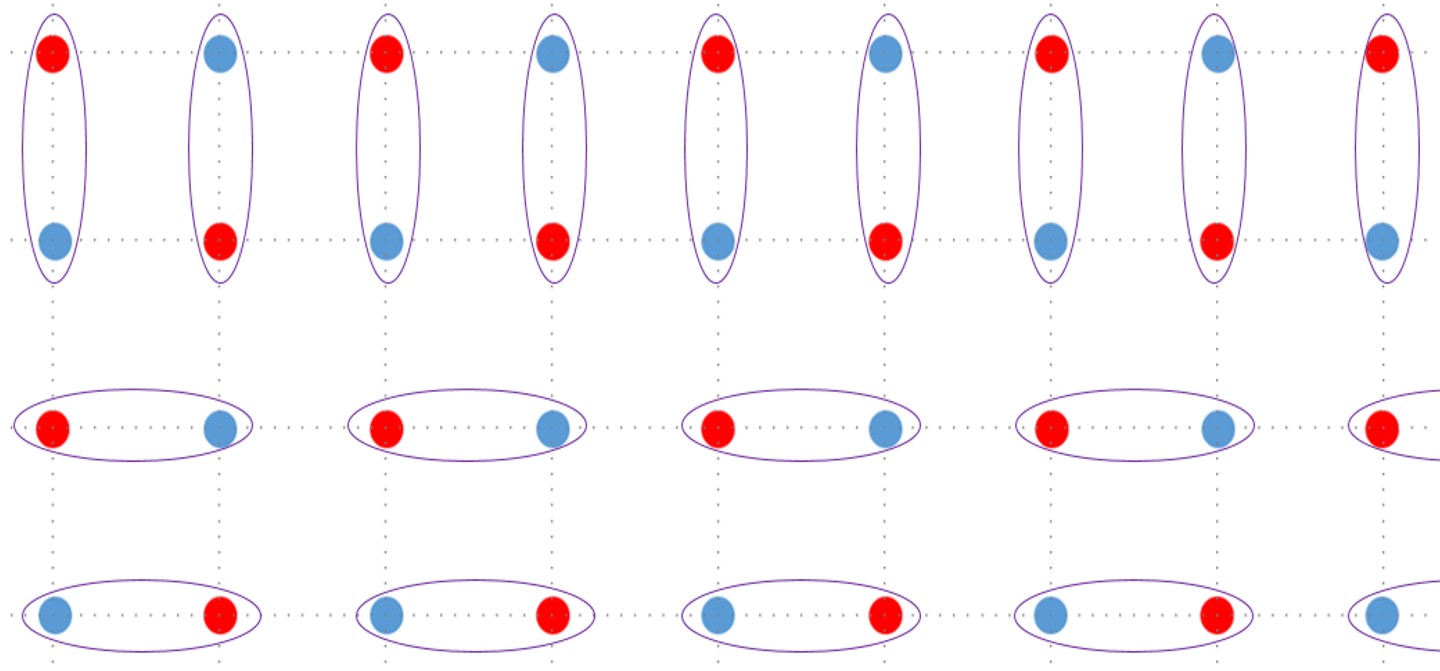}
 \caption{When the dipoles are densely packed, the notion of separate species of dipoles breaks down.  As seen in the top row, this configuration of charges could be regarded as closely packed $y$-oriented dipoles.  But, as seen in the bottom row, the same configuration of charge could be regarded as closely packed $x$-oriented dipoles.  If this charge configuration were set in motion, there would be a fundamental ambiguity in defining a ``dipole current."  This ambiguity is connected with the fact that the current tensor of this theory must be a symmetric tensor.}
 \label{fig:dipole}
 \end{figure*}

\subsection{Currents and the Biot-Savart Law}

In addition to static charge distributions, we should also think about how to handle steady current flows.  The fundamental charges are fractons, which cannot freely hop and therefore have no sense of current.  The natural mobile objects of the theory are the dipolar bound states.  In order to keep track of such dipole motion, one's first instinct might be to define a (non-symmetric) current tensor $\mathcal{J}_{ij}$ representing the current of the $i$ directed dipoles in the $j$ direction.  However, there is a fundamental ambiguity in this definition.  For example, consider the close-packed charge configuration in Figure \ref{fig:dipole}.  There is not a unique way of defining either the dipole density or $\mathcal{J}_{ij}$ in such a case.  Microscopically, an operator hopping an $i$ dipole in the $j$ direction is the same operator hopping a $j$ dipole in the $i$ direction, so the true microscopic current operator is actually a symmetric tensor $J_{ij}$.  This can further be seen by noting that a source term for the gauge field in the Hamiltonian, $A^{ij}J_{ij}$, would not be sensitive to any antisymmetric components.  We shall comment further on the meaning and fundamental definition of the microscopic current $J_{ij}$ shortly.

In terms of the microscopic current, the Hamiltonian takes the form:
\begin{equation}
\begin{split}
\int\bigg(\frac{1}{2}E^{ij}E_{ij} + \frac{1}{2}B^{ij}B_{ij} + A^{ij}J_{ij}\bigg) = \\
\int\bigg(\frac{1}{2}E^{ij}E_{ij} + \frac{1}{2} A^{ij}\epsilon^{\,\,\,ab}_{i}\partial_a B_{bj} + A^{ij}J_{ij}\bigg)
\end{split}
\end{equation}
Noting that $A_{ij}$ is a symmetric tensor, we evaluate the equation of motion for $E_{ij}$ to find the following generalized Ampere's equation relating the current to the fields:
\begin{equation}
\frac{1}{2}(\epsilon^{iab}\partial_a B_b^{\,\,\,j} + \epsilon^{jab}\partial_a B_b^{\,\,\,i}) = -J^{ij} - \partial_t E^{ij}
\end{equation}
We note that this generalized Ampere's equation can actually be used as the fundamental \emph{definition} of the microscopic current $J_{ij}$.  Just as the charge $\rho$ can be defined as the violations of $\partial_i\partial_j E^{ij} = 0$ from the ``pure" gauge theory, $J_{ij}$ essentially represents the deviations of Ampere's equation from the pure gauge theory.  In order to make connection between this definition and the concept of rates of charge hopping processes, we can use this equation to derive a continuity equation relating the charge and current.  Applying $\partial_i\partial_j$ to each side of Ampere's equation, we obtain:
\begin{equation}
\partial_t\rho + \partial_i\partial_j J^{ij} = 0
\end{equation}
as the generalized continuity equation.  The intuition of this equation is that $J^{ij}$ inherently represents the rate of \emph{multi-body} hopping processes, instead of the single-body motion captured by a more conventional current vector.  For example, dipole motion is one sort of process contributing to $J^{ij}$, as expected.  But other multi-body processes, such as a fracton hopping while emitting an extra dipole, also contribute.  This intuition can be directly confirmed in the lattice models \cite{alex,cenke1,cenke2}, where all such multi-body processes act as sources for Ampere's equation.

To have a steady current, our continuity equation demands $\partial_i\partial_j J^{ij} = 0$.  Assuming this to be the case, so that there are steady currents and electric fields, we can drop the $\partial_t E^{ij}$ term.  We can then rearrange Ampere's equation as:
\begin{equation}
\partial_a(\epsilon^{abi}B_b^{\,\,\,j} + \epsilon^{abj}B_b^{\,\,\,i}) = -2J^{ij}
\label{fara}
\end{equation}
We can obtain a general solution for the quantity in parentheses as:
\begin{equation}
\begin{split}
\epsilon^{abi}B_b^{\,\,\,j} + \epsilon^{abj}B_b^{\,\,\,i} =& \\
\bigg(-\frac{1}{2\pi}\int dr' J^{ij}(r')& \frac{(r-r')^a}{|r-r'|^3}\bigg) + \epsilon^{a\ell k}\partial_\ell\lambda_k^{\,\,\,ij}
\end{split}
\end{equation}
for arbitrary tensor $\lambda_{kij}$.  Applying $\epsilon_{adi}$ to this equation yields:
\begin{equation}
\begin{split}
&3B_{d}^{\,\,\,j} - B^i_{\,\,\,i}\delta_{d}^{\,\,\,j} = \\
&\bigg(-\frac{1}{2\pi}\int dr' J^{ij}(r')\epsilon_{adi}\frac{(r-r')^a}{|r-r'|^3}\bigg) + \partial_d\lambda_i^{\,\,\,ij} -\partial_i\lambda_{d}^{\,\,\,ij}
\end{split}
\end{equation}
By its definition, we note that $B$ is traceless, $B^i_{\,\,\,i} = 0$, so we are left with:
\begin{equation}
\begin{split}
B^{ij} = \bigg(-\frac{1}{6\pi}\int dr' J^{j}_{\,\,\,k}(r')\epsilon^{ik\ell}\frac{(r-r')_\ell}{|r-r'|^3}\bigg) + \\
\frac{1}{3}(\partial^i\lambda_k^{\,\,\,kj} -\partial_k\lambda^{ikj})
\label{general}
\end{split}
\end{equation}
This is the generic solution to our generalized Ampere's equation.  However, we must now pick the solution that also obeys the absence of magnetic charges, $\partial_iB^{ij} = 0$.  By good fortune, we note that the choice $\lambda_{kij} = 0$ is the solution which obeys this property.  The final result for the magnetic field generated by a steady current is:
\begin{equation}
B^{ij} = -\frac{1}{6\pi}\int dr' J^{j}_{\,\,\,k}(r')\epsilon^{ik\ell}\frac{(r-r')_\ell}{|r-r'|^3}
\label{biot}
\end{equation}
This equation serves as the generalized Biot-Savart law for this theory.  For an arbitrary steady current configuration, we can use this equation to calculate the resulting magnetic field tensor.  Note that the Biot-Savart law obeys the same scaling as in conventional electromagnetism.  Thus, for example, the magnetic field of a current-carrying wire will fall off as $1/r$ with distance $r$ away from the wire, just like a conventional current.  Whereas electric fields of static charges were abnormally energetically costly in this theory, magnetic fields of steady currents are much more in line with standard electromagnetism.

\subsection{Magnetic Particles}

The theory considered in this section, alone of the rank 2 theories, is not self-dual.  The electric particles of this theory are scalars, but the magnetic charges are vectors, $\partial_i B^{ij} = \tilde{\rho}^j$, which behave as 2-dimensional particles.  Whereas the fields associated with the magnetic particles could be easily obtained for a self-dual theory, for this theory the magnetic results are not automatic.  The calculation of these fields is similar to other calculations in this paper and would distract from the main line of development, so we relegate the calculations to Appendix A.  We here simply state the results.  The magnetic field corresponding to a magnetic charge $\partial_i B^{ij} = p^j\delta^{(3)}(r)$ is:
\begin{equation}
B^{ij} = -\frac{1}{16\pi}\bigg(-5\frac{p^jr^i}{r^3} - \frac{p^ir^j}{r^3} + \frac{(p\cdot r)\delta^{ij}}{r^3} + 3\frac{(p\cdot r)r^ir^j}{r^5}\bigg)
\end{equation}
The Lorentz force on a magnetic particle is:
\begin{equation}
F^i = p^j(P^{ik}B_{kj} - \epsilon^{ik\ell}v_k\hat{p}_\ell\hat{p}^n E_{jn})
\end{equation}
where $P^{ik}$ is the projector into the plane transverse to $p^j$.  The current of magnetic particles takes the form of a traceless non-symmetric tensor $\tilde{J}_{ij}$, obeying a continuity equation:
\begin{equation}
\partial_t \tilde{\rho}^j + \partial_i \tilde{J}^{ij} = 0
\end{equation}
The generalized Faraday's equation is:
\begin{equation}
\epsilon^{iab}\partial_aE_b^{\,\,\,j} = \partial_tB^{ij} + \tilde{J}^{ij}
\end{equation}
For a steady magnetic current configuration, the dual Biot-Savart law is:
\begin{equation}
E^{ij} = \frac{1}{8\pi}\int dr' (\tilde{J}_k^{\,\,\,j}(r')\epsilon^{k\ell i} + \tilde{J}_k^{\,\,\,i}(r')\epsilon^{k\ell j})\frac{(r-r')_\ell}{|r-r'|^3}
\end{equation}

\subsection{Summary of Maxwell Equations}

Our task complete, we now take a moment to collect the generalized Maxwell equations, from which all of the other results follow.  For the scalar charge theory, these equations take the form:
\begin{equation}
\boxed{
\begin{split}
\partial_i&\partial_j E^{ij} = \rho \\
&\partial_i B^{ij} = \tilde{\rho}^j \\
\epsilon^{iab}\partial_a &E_b^{\,\,\,j} = \partial_t B^{ij} + \tilde{J}^{ij}\\
\frac{1}{2}(\epsilon^{iab}\partial_a B_b^{\,\,\,j} + \epsilon^{jab}&\partial_a B_b^{\,\,\,i}) = -\partial_t E^{ij} - J^{ij}
\end{split}
}
\end{equation}
where $\rho$ and $J^{ij}$ are the charge and current of electric particles, and $\tilde{\rho}^j$ and $\tilde{J}^{ij}$ are the charge and current of magnetic particles.

\section{Traceless Scalar Charge Theory}

\subsection{Electrostatic Fields}

Let us now consider a different rank 2 theory, where we will take the same Gauss's law, $\partial_i\partial_j E^{ij} = \rho$, but will also impose the condition that the electric field tensor is traceless, $E^i_{\,\,\,i} = 0$.  As discussed in Reference \onlinecite{me}, we are free to impose this constraint identically on the entire Hilbert space, without charges.  One could take trace charges into account, but the analysis would become more tedious, without qualitatively affecting the results, so we shall avoid such a discussion here.  As before, the fundamental charges in this theory are fractonic, totally unable to move.  However, the dipolar bound states, which were formerly fully mobile, are now 2-dimensional particles in this theory, since they can only hop transversely while obeying the conservation laws.

Once again, we wish to first calculate the electric field of an isolated point charge.  Particle independence and rotational symmetry dictate that the electric field tensor must take the form of Equation \ref{generalcoul}.  In order to satisfy Gauss's law, we must have $\beta = \alpha + 1/4\pi$, as before.  However, the magnetostatic constraint is different in this case.  For this theory, the appropriate magnetic tensor is the symmetric tensor $B^{ij} = \frac{1}{2}(\epsilon^{iab}\partial_a A_b^{\,\,\,j} + \epsilon^{jab}\partial_a A_b^{\,\,\,i})$.  The magnetostatic condition is then $\partial_t B^{ij} = \frac{1}{2}(\epsilon^{iab}\partial_a E_b^{\,\,\,j} + \epsilon^{jab}\partial_a E_b^{\,\,\,i}) = 0$.  However, in this case, we find that this constraint is no constraint at all.  The electric field tensor of Equation \ref{generalcoul} will satisfy this constraint for any values of $\alpha$ and $\beta$, so this constraint is of no use to us.  In order to fully determine the electric field, we must then resort to the tracelessness condition, $E^i_{\,\,\,i} = 0$, which gives the following:
\begin{equation}
E^i_{\,\,\,i} = q\bigg(\alpha \frac{\delta^i_i}{r} + \beta \frac{r^ir_i}{r^3}\bigg) = q\frac{3\alpha+\beta}{r} = 0
\end{equation}
Our two conditions are then $3\alpha + \beta = 0$ and $\beta = \alpha + 1/4\pi$, which has the solution $\alpha = -\frac{1}{16\pi}$ and $\beta = \frac{3}{16\pi}$.  The electric field of an electric monopole for this theory is then:
\begin{equation}
E^{ij}_{mono} = \frac{q}{16\pi}\bigg(\frac{3r^ir^j}{r^3} - \frac{\delta^{ij}}{r}\bigg)
\end{equation}
The corresponding dipole field is $-p^k\partial_k E_{q=1}^{ij}$, which yields:
\begin{equation}
E^{ij}_{dipole} = -\frac{1}{16\pi}\bigg(\frac{\delta^{ij}(r\cdot p)}{r^3} + 3\frac{(p^ir^j + r^ip_j)}{r^3} - 9 \frac{r^ir^j(r\cdot p)}{r^5}\bigg)
\end{equation}

\subsection{Potential Formulation}

Once again, it is desirable to obtain a potential formulation for this theory.  However, the magnetostatic constraint, $\frac{1}{2}(\epsilon^{iab}\partial_a E_b^{\,\,\,j} + \epsilon^{jab}\partial_a E_b^{\,\,\,i}) = 0$, is a bit more complicated in this case, and it is not obvious at first glance what potential formulation to construct.  However, we can take our inspiration from the previous case and start with $\partial_i\partial_j \phi$.  In general, this is not a traceless tensor, so we remove the trace component and try a potential ansatz of the form:
\begin{equation}
E_{ij} = \partial_i\partial_j\phi -\frac{1}{3}\delta_{ij} \partial^2 \phi
\label{potential}
\end{equation}
If this potential formulation is to work, then by the same rotational symmetry arguments as before, the form $\phi = Cr$ for some constant $C$ must yield the correct field upon solving Gauss's law.  This equation gives $\partial_i\partial_j E^{ij} = \frac{2}{3}(\partial^2)^2 \phi = -\frac{2}{3}8\pi C \delta^{(3)}(r) = q\delta^{(3)}(r)$, so we require $C = -3q/16\pi$.  Upon taking the appropriate derivatives, we find that this potential yields the correct electric field of a point charge.  While we have not derived the form of the potential directly from the magnetostatic condition, this is actually not necessary to demonstrate its correctness.  This potential formulation (obtained via educated guess) works for the point charge.  Then, by linearity of all equations involved, we can superpose potentials to get the correct electric field for an arbitrary electrostatic configuration of charges.  Thus, the potential formulation of Equation \ref{potential} is rigorously correct for all electrostatic problems.  The potentials for the electric monopole and dipole are given as follows:
\begin{equation}
\phi_{mono} = -\frac{3qr}{16\pi}\,\,\,\,\,\,\,\,\,\,\,\,\,\,\,\,\,\phi_{dip} = \frac{3(p\cdot r)}{16\pi r}
\end{equation}
As before, the potential is of direct physical significance.  The energy stored in the electric field of an arbitrary electrostatic configuration is given by:
\begin{equation}
\begin{split}
\epsilon = \frac{1}{2}\int E^{ij}E_{ij} = &\frac{1}{2}\int E^{ij}(\partial_i\partial_j\phi -\frac{1}{3}\delta_{ij}\partial^2\phi) = \\
&\frac{1}{2}\int \partial_i\partial_j E^{ij}\phi = \frac{1}{2}\int\rho\phi
\end{split}
\end{equation}
where we have made use of the tracelessness condition.  As in the previous case, we see that the potential is a direct measure of the energy associated with a particle being at a particular location.  Once again, there are no ``self-energy" contributions, and the factor of $1/2$ serves to eliminate double counting of particle pairs.  Note that the coefficient of the monopole potential in this case, $3/16\pi$, is larger than that in the previous traceful case, $1/8\pi$.  Therefore, the traceless theory has stronger interactions between particles than its traceful cousin.

\subsection{Lorentz Force}

As in the traceful theory, the fundamental particles are fractons and cannot respond directly to the electromagnetic fields.  However, we still have mobile dipolar bound states, which in this theory are 2-dimensional particles, hopping only in the transverse direction.  The phases picked up upon hopping will still be of the form $-p_j A^{ji}$, but the longitudinal component will not be felt, so we should project into the transverse plane, obtaining the effective vector potential as:
\begin{equation}
A_{eff}^i = -P^{ik}\hat{p}^j A_{jk}
\end{equation}
where we have defined the projection operator $P^{ik}=(\delta^{ik}-\hat{p}^i\hat{p}^k)$, which projects onto the transverse plane.  The effective magnetic field is then the out-of-plane component of the curl of this vector potential:
\begin{equation}
\begin{split}
B_{eff}^i = \hat{p}^i\hat{p}^j\epsilon^{jk\ell}\partial_k A^{eff}_\ell &= -\hat{p}^i\hat{p}^j\hat{p}^m \epsilon_j^{\,\,\,k\ell}\partial_k  A_{\ell m} = \\
&-\hat{p}^i\hat{p}^j\hat{p}^k B_{jk}
\end{split}
\end{equation}
The corresponding effective electric field is $E_{eff}^i = \partial_tA_{eff}^i = -P^{ik}\hat{p}^j E_{jk}$, which lies in the plane, as appropriate.  The generalized Lorentz force on a dipole $p$ in this theory is then:
\begin{equation}
F^i = -p^j (P^{ik}E_{jk} + \epsilon^{ik\ell}v_k \hat{p}_\ell\hat{p}^nB_{jn})
\end{equation}
Note that, since the velocity and the effective $E$ lie in the plane, and the effective $B$ is perpendicular to the plane, the Lorentz force always lies in the plane, consistent with the 2-dimensional nature of the dipoles.

Let us now take the electromagnetic fields to be generated by some electrostatic configuration of charges, so that $B$ vanishes and we may use our earlier potential formulation.  The force law then simplifies to:
\begin{equation}
F^i = -p^j P^{ik}(\partial_j\partial_k \phi - \frac{1}{3}\delta_{jk} \partial^2 \phi)
\end{equation}
The work done to move a dipole against the field from point 1 to point 2 in the plane of its motion is then the line integral of the force against the field.  Since the path lies in the plane, we may drop the projection operator in the above equation, as the transverse component will be picked out anyway:
\begin{equation}
W = -\int_1^2 dx_i F^i = \int_1^2 dx^i p^j (\partial_j\partial_i \phi - \frac{1}{3}\delta_{ji}\partial^2 \phi)
\end{equation}
The second term will not contribute, since $p^i dx_i = 0$ for motion in the plane, leaving us with:
\begin{equation}
W = \int_1^2 dx^i\partial_i (p^j\partial_j\phi) = (p^j \partial_j\phi)_2 - (p^j\partial_j\phi)_1
\end{equation}
We see that, once again, the potential energy of a dipole $p$ is given by $V = p^j\partial_j\phi$, as it should be, based on the earlier discussion of the potential.

In order to get a fully mobile charge in this theory, one must consider a bound state which is not only neutral, but also has no net dipole.  Such quadrupolar bound states would couple only weakly to the gauge field, via derivatives of $A$ instead of $A$ itself.  Furthermore, not all such quadrupolar bound states are stable.  Certain quadrupoles can decay directly to the vacuum, releasing their energy into the gapless gauge mode.  The only stable quadrupoles are those with a nonzero value of $\int \rho x^2$, which prevents decay by the quadrupolar conservation law.  We shall not further investigate the properties of such quadrupolar bound states.

\subsection{Currents and the Biot-Savart Law}

We also wish to characterize steady current distributions in this theory.  As in the traceful theory, the fundamental microscopic current tensor is a symmetric tensor $J_{ij}$.  However, in this case $J_{ij}$ must be a traceless tensor, $J^i_{\,\,\,i} = 0$, since the trace component represents the rate of processes which violate the trace constraint on the electric field.  In terms of this current, the Hamiltonian takes the form:
\begin{equation}
\begin{split}
\int\bigg(\frac{1}{2}&E^{ij}E_{ij} + \frac{1}{4}(\epsilon^{iab}\partial_aA_b^{\,\,\,j}+\epsilon^{jab}\partial_aA_b^{\,\,\,i})B_{ij} + A^{ij}J_{ij}\bigg) = \\
&\int\bigg(\frac{1}{2}E^{ij}E_{ij} + \frac{1}{2}A^{ij}\epsilon_i^{\,\,ab}\partial_aB_{bj} + A^{ij}J_{ij}\bigg)
\end{split}
\end{equation}
The corresponding Ampere's equation for the time evolution of $E_{ij}$ is:
\begin{equation}
\frac{1}{2}(\epsilon_i^{\,\,ab}\partial_a B_{bj} + \epsilon_j^{\,\,ab}\partial_a B_{bi}) = -J_{ij} - \partial_t E^{ij}
\end{equation}
Once again, this current tensor will obey a continuity equation:
\begin{equation}
\partial_t\rho + \partial_i\partial_j J^{ij} = 0
\end{equation}
so a steady current requires $\partial_i\partial_j J^{ij} = 0$.  For steady currents, we drop the electric field term in the Ampere's equation.  Ampere's equation is actually formally the same as that obtained in the traceful theory, so we expect the Biot-Savart laws to be almost identical.  The only differences are that $B_{ij}$ is now constrained to be symmetric, and the condition for the absence of magnetic charge is different ($\partial_i\partial_j B^{ij} = 0$ in the present case, versus $\partial_i B^{ij} = 0$ in the previous case).  Nevertheless, the generic form of Equation \ref{general}, in terms of arbitrary tensor $\lambda_{kij}$, is still valid.  Noting that the final result must be symmetric, we guess the form:
\begin{equation}
\lambda_{kij} = -\frac{1}{2\pi}\int dr' J^k_{\,\,\,\ell}(r')\epsilon^{j\ell i}\frac{1}{|r-r'|}
\end{equation}
This is automatically a solution of the generalized Ampere's equation.  The resulting magnetic field tensor is:
\begin{equation}
B^{ij} = -\frac{1}{6\pi}\int dr' (J^j_{\,\,\,k}(r')\epsilon^{ik\ell} + J^i_{\,\,\,k}(r')\epsilon^{jk\ell})
\frac{(r-r')_\ell}{|r-r'|^3}
\end{equation}
which is manifestly symmetric.  It also satisfies the equation $\partial_i\partial_j B^{ij} = 0$.  This is therefore the correct generalized Biot-Savart law for this theory.  From this equation, we can construct the magnetic field tensor for an arbitrary steady current configuration.

\subsection{Summary of Maxwell Equations}

The generalized Maxwell equations for the traceless scalar charge theory take the form:
\begin{equation}
\boxed{
\begin{split}
&\partial_i\partial_jE^{ij} = \rho \\
&\partial_i\partial_jB^{ij} = \tilde{\rho} \\
\frac{1}{2}(\epsilon_{iab}\partial^a E^b_{\,\,\,j} +& \epsilon_{jab}\partial^a E^b_{\,\,\,i}) = \partial_t B_{ij} + \tilde{J}_{ij} \\
\frac{1}{2}(\epsilon_i^{\,\,ab}\partial_a B_{bj} +& \epsilon_j^{\,\,ab}\partial_a B_{bi}) =  - \partial_t E^{ij} - J_{ij} \\
&E^i_{\,\,\,i} = 0 \\
&B^i_{\,\,\,i} = 0
\end{split}}
\end{equation}
We have included the tracelessness constraint alongside the other Maxwell equations.  But note that the time evolution equation for $E$ ensures that, as long as the initial condition for $E$ is traceless, it will automatically remain traceless under time evolution.  We have also written out the tracelessness of $B$ for symmetry purposes.  Note that, unlike the previous theory, the Maxwell equations here have a nice symmetry between electric and magnetic quantities, reflecting the self-duality of the theory.

\section{Vector Charge Theory}

\subsection{Electrostatic Fields}

We will now switch gears and move to a theory with a different Gauss's law altogether.  Our degrees of freedom will still be that of a rank 2 symmetric $U(1)$ tensor $A_{ij}$, but we will now take our Gauss's law to be $\partial_i E^{ij} = \rho^j$, with vector charge $\rho^j$.  An isolated charge will provide a delta function source for this Gauss's law:
\begin{equation}
\partial_i E^{ij} = p^j\delta^{(3)}(r)
\end{equation}
for some charge vector $p^j$.  The solution to this equation must be a symmetric rank 2 tensor, must depend only on $r^j$ and $p^j$, must be linear in $p^j$, and by dimensional analysis must scale as $1/r^2$.  The possible terms are then:
\begin{equation}
\begin{split}
E^{ij} = \alpha\frac{(p^ir^j+ r^ip^j)}{r^3} + \beta \frac{(p\cdot r)\delta^{ij}}{r^3} + \gamma \frac{(p\cdot r)r^ir^j}{r^5} + \\
\mu \frac{(\epsilon^{ik\ell}r_kp_\ell r^j + \epsilon^{jk\ell}r_k p_\ell r^i)}{r^4}
\end{split}
\label{1dgeneral}
\end{equation}
We can then take a derivative, being careful about delta functions at the origin (making use of some useful formulas from Appendix B).  The result is:
\begin{equation}
\begin{split}
\partial_i E^{ij} = (\alpha+\beta) \bigg(\frac{p^j}{r^3} - \frac{3(p\cdot r)r^j}{r^5}\bigg) + \mu\bigg(\frac{\epsilon^{jk\ell}r_kp_\ell}{r^4}\bigg) +\\
\frac{4\pi}{3}(4\alpha + \beta + \gamma) p^j \delta^{(3)}(r)
\label{div}
\end{split}
\end{equation}
In order to solve Gauss's law, we therefore need $\beta = -\alpha$, $\mu = 0$, and $\gamma = \frac{3}{4\pi} - 3\alpha$.  This takes us from four unknown coefficients down to one, $\alpha$.  We must then resort to the magnetostatic condition.  For this theory, the magnetic field tensor is given by $B_{ij} = \epsilon_{iab}\epsilon_{jcd}\partial^a\partial^c A^{bd}$.  The magnetostatic condition on the electric field is then $\epsilon_{iab}\epsilon_{jcd}\partial^a\partial^c E^{bd} = 0$.  To start, let us simply look at the trace component of this constraint:
\begin{equation}
\begin{split}
\epsilon_{iab}\epsilon_{icd}\partial^a &\partial^c E^{bd} = \partial^2 E^i_{\,\,\,i} - \partial_i\partial_j E^{ij} =\\
&\partial^2 E^i_{\,\,\,i} - \partial_i \rho^i = 0
\label{trace}
\end{split}
\end{equation}
where we have made use of the Gauss's law.  Taking the trace of our general formula, Equation \ref{1dgeneral}, yields $E^i_{\,\,\,i} = (2\alpha + 3\beta + \gamma)(p\cdot r)/r^3$, which (up to a constant) is formally equivalent to the potential energy of an ordinary electromagnetic dipole.  We can then take $\partial^2$ of this quantity by appealing to the ordinary Poisson equation and the charge distribution of an ordinary dipole.  (Directly differentiating this potential is actually quite subtle, for distributional reasons, but a direct calculation yields the same results \cite{dipole}.)  The result is:
\begin{equation}
\partial^2 E^i_{\,\,\,i} = 4\pi(2\alpha+3\beta+\gamma)p^j \partial_j\delta^{(3)}(r)
\end{equation}
In order to satisfy Equation \ref{trace}, we must therefore have $(2\alpha+3\beta+\gamma) = 1/4\pi$.  When combined with our earlier results, $\beta = -\alpha$ and $\gamma = \frac{3}{4\pi} - 3\alpha$, we obtain $\alpha = \frac{1}{8\pi}$, $\beta = -\frac{1}{8\pi}$, $\gamma = \frac{3}{8\pi}$.  We have now checked that this form satisfies the trace magnetostatic condition, but it can also be verified that the full magnetostatic condition is satisfied.  The electrostatic field for a point charge $p^j$ then takes the final form:
\begin{equation}
E^{ij} = \frac{1}{8\pi}\bigg(\frac{(p^ir^j+r^ip^j)}{r^3} - \frac{(p\cdot r)\delta^{ij}}{r^3} + 3\frac{(p\cdot r)r^ir^j}{r^5}\bigg)
\end{equation}

\subsection{Potential Formulation}

Once again, it will be advantageous to seek a potential formulation for the theory.  From our magnetostatic condition, $\epsilon_{iab}\epsilon_{jcd}\partial^a \partial^c E^{bd}$, we can see that either the first or second index of $E^{ij}$ should have a derivative in it, so $E^{ij}$ should have the form $E^{ij} = \partial_i\phi_j +\partial_j\psi_i$ for vectors $\phi_i$ and $\psi_i$.  In order to satisfy index symmetry, we set these vectors to be equal.  We also add in a factor of $-1/2$ for later convenience, writing the electric field as:
\begin{equation}
E_{ij} = -\frac{1}{2}(\partial_i \phi_j + \partial_j \phi_i)
\end{equation}
The most general form for $\phi^i$ is:
\begin{equation}
\phi^j = \alpha' \frac{(p\cdot r)r^j}{r^3} + \beta' \frac{p^j}{r} + \gamma' \frac{\epsilon^{jk\ell}p_kr_\ell}{r^2}
\label{vecpot}
\end{equation}
Taking derivatives yields:
\begin{equation}
\begin{split}
E^{ij} = \frac{(\beta'-\alpha')}{2}\frac{(p^ir^j+r^ip^j)}{r^3} - \alpha' \frac{(p\cdot r)\delta^{ij}}{r^3} +\\
3\alpha'\frac{(p\cdot r)r^ir^j}{r^5} + \gamma'\frac{(\epsilon^{jk\ell}p_kr_\ell r^i + \epsilon^{ik\ell}p_kr_\ell r^j)}{r^4}
\end{split}
\end{equation}
The magnetostatic condition is automatic, but we must check the Gauss's law.  Following the logic of Equation \ref{div}, we obtain that $\gamma' = 0$, $\alpha' = 1/8\pi$, and $\beta' = 3/8\pi$, which correctly reproduces the desired electric field for a point charge.  The final result for the potential of a point charge is:
\begin{equation}
\phi^j = \frac{1}{8\pi}\bigg(\frac{(p\cdot r)r^j}{r^3} + 3\frac{p^j}{r}\bigg)
\end{equation}
Note that, unlike the previous two cases, there is no scalar potential formulation, but rather a vector potential.  (It can be explicitly checked that no derivatives of any scalar potential can solve the Gauss's law.)  This makes some intuitive sense, since now the Gauss's law is a three-component equation, corresponding to three particle degrees of freedom.  Any potential formulation must at least capture these three degrees of freedom, so a vector potential is the best we can do.  But this is still a significant simplification over the original tensor formulation, and the potential of an arbitrary charge distribution can be built up by superposing the point charge potential given above.

In order to give a physical interpretation to the potential, let us look at the energy stored in the electric field of a static charge configuration:
\begin{equation}
\begin{split}
\epsilon = \frac{1}{2}\int E^{ij}E_{ij} = -\frac{1}{4}\int E^{ij}(\partial_i\phi_j + \partial_j \phi_i) = \\
\frac{1}{2}\int \partial_i E^{ij}\phi_j = \frac{1}{2}\int \rho^j\phi_j
\end{split}
\end{equation}
which is very similar in fashion to the previous sections.  The potential $\phi^j$ represents the potential energy of charge species $p^j$.  One key difference from the previous theories is that the potential $\phi^j$ of a point charge blows up at the charge's location, like in conventional electromagnetism, so there are ``self-energy" contributions to the integral which should be handled with care.

\subsection{Lorentz Force}

A vector point charge $p^j$ can only hop along the $\hat{p}^j$ direction.  The phases picked up upon completing such hops are $\hat{p}_i\hat{p}_j A^{ij}$.  Therefore, these 1-dimensional particles only feel a one-dimensional effective electric field, given by $E_{eff}^i = (\hat{p}_j\hat{p}_k E^{jk})\hat{p}^i$.  These particles do not feel any effects from the magnetic field, since their one-dimensional trajectories cannot enclose any flux.  The Lorentz force law takes the particularly simple form:
\begin{equation}
F^i = (\hat{p}_j\hat{p}_kE^{jk})p^i
\end{equation}
For an electrostatic configuration, $E^{ij} = -\frac{1}{2}(\partial^i\phi^j + \partial^j\phi^i)$, we have $F^i = -(\hat{p}_j\hat{p}_k\partial^j\phi^k)p^i$.  The effective force is just the projection of $-p_k\partial^i\phi^k$ along the $\hat{p}^i$ direction.  We now wish to calculate the work done in moving a particle along the line of its motion against the force of a field.  For this purpose, we are free to use the unprojected force, since the other components will not contribute anyway:
\begin{equation}
W = -\int_1^2 dx_i F^i = \int_1^2 dx_i \partial^i(p_k\phi^k) = (p_k\phi^k)_2 - (p_k\phi^k)_1
\end{equation}
We therefore have that the potential energy for charge $p_j$ is given by $V = p_j\phi^j$, as we already found.

In addition to the 1-dimensional particles, one can construct a fully mobile, yet still topologically nontrivial excitation by looking at a bound state with zero net charge $p^j$ but a nonzero charge angular moment.  This bound state is fully mobile, yet cannot be created locally (due to its charge angular moment) and is therefore stable against decay into the vacuum.  We refer to such a nontrivial excitation, carrying zero charge but nonzero charge angular moment, as a ``chiron," because all the good names were taken.  In the lattice models \cite{alex,cenke1,cenke2}, one can verify that the phase picked up by a chiron carrying angular moment $L^j$ hopping in the $i$ direction is $\epsilon^{jk\ell}L_j\partial_k A_{\ell i}$.  This quantity then serves as the effective chiron vector potential, $A^{eff}_i = \epsilon^{jk\ell}L_j\partial_k A_{\ell i}$.  The effective chiron magnetic field is $B_{eff}^i = \epsilon^{ijk} \epsilon^{\ell mn}L_\ell \partial_j\partial_m A_{nk} = L_jB^{ij}$.  The Lorentz force on a chiron is therefore:
\begin{equation}
F^i = L^j(\epsilon_{jk\ell}\partial^kE^{\ell i} + \epsilon^{ik\ell}v_k B_{\ell j})
\end{equation}
Note that the chirons will respond to a uniform magnetic field, but only to derivatives of the electric field.  In a sense, the 1-dimensional particles represent the fundamental unit of response to a uniform electric field, while the chirons represent the fundamental unit of response to a uniform magnetic field.

\subsection{Currents and the Biot-Savart Law}

As in the previous theories, the microscopic current operator in this theory will be a symmetric tensor $J_{ij}$ representing the rate of allowed hopping processes, including free longitudinal motion and also multi-body transverse motion.  In terms of the current operator, the Hamiltonian of the theory is given by:
\begin{equation}
\begin{split}
\int\bigg(&\frac{1}{2}E^{ij}E_{ij} + \frac{1}{2}\epsilon^{iab}\epsilon^{jcd}\partial_a\partial_cA_{bd}B_{ij} + A^{ij}J_{ij}\bigg) = \\
\int&\bigg(\frac{1}{2}E^{ij}E_{ij} + \frac{1}{2}A_{ij}\epsilon^{iab}\epsilon^{jcd}\partial_a\partial_cB_{bd} + A^{ij}J_{ij}\bigg)
\end{split}
\end{equation}
The time evolution equation for the electric field is then:
\begin{equation}
\epsilon^{iab}\epsilon^{jcd}\partial_a\partial_cB_{bd} = -J^{ij} -\partial_t E^{ij}
\end{equation}
Taking a derivative, we find that the charge and current are related by the continuity equation:
\begin{equation}
\partial_t \rho^j +\partial_i J^{ij} = 0
\end{equation}
so a steady current configuration requires $\partial_iJ^{ij} = 0$.  Taking such a steady current configuration, we can rewrite our Ampere's equation as:
\begin{equation}
\partial_a\partial_c(\epsilon^{iab}\epsilon^{jcd}B_{bd}) = -J^{ij}
\end{equation}
From our earlier work on the electric field corresponding to $\partial_i\partial_jE^{ij} = \delta^{(3)}(r)$, we can write a generic solution as:
\begin{equation}
\begin{split}
&\epsilon^{iab}\epsilon^{jcd}B_{bd} = \\
-&\int dr'J^{ij}(r')\bigg(\alpha\frac{\delta^{ac}}{|r-r'|} +\beta\frac{(r-r')^a(r-r')^c}{|r-r'|^3}\bigg)
\end{split}
\end{equation}
subject to the constraint $4\pi(\beta - \alpha) = 1$.  (We could also have added in a solution to the homogeneous equation, but it will turn out that we do not need it in this case.)  Applying $\frac{1}{4}\epsilon^{iak}\epsilon^{jc\ell}$ to both sides of the equation above and relabeling some indices yields:
\begin{equation}
\begin{split}
&B^{ij} = \\
-&\int dr' J^{k\ell}(r')\epsilon^{ika}\epsilon^{j\ell c}\bigg(\alpha\frac{\delta^{ac}}{|r-r'|} +\beta\frac{(r-r')^a(r-r')^c}{|r-r'|^3}\bigg)
\end{split}
\end{equation}
We also need the constraint that $\partial_iB^{ij} = 0$ (and equivalently, $\partial_jB^{ij} = 0$).  This can be easily taken care of if the quantity in parentheses has the form $\partial^a\partial^c\phi$ for some scalar $\phi$.  This means that the appropriate choice of $\alpha$ and $\beta$ are exactly those appropriate to the electric field of a point charge in the traceful scalar charge theory.  The resulting magnetic field tensor is:
\begin{equation}
\begin{split}
&B^{ij} = \\
&\frac{1}{8\pi}\int dr' J^{k\ell}(r')\epsilon^{ika}\epsilon^{j\ell c}\bigg(\frac{\delta^{ac}}{|r-r'|} -\frac{(r-r')^a(r-r')^c}{|r-r'|^3}\bigg)
\end{split}
\end{equation}
The above equation serves as the generalized Biot-Savart law for this theory and allows us to construct the magnetic field for an arbitrary steady current distribution.  Note that this Biot-Savart law is stronger by one power than the conventional electromagnetic one.  The integrand falls off as $1/r$ instead of $1/r^2$.  This will cause currents in this theory to be extremely energetically costly.  For example, whereas the field of a current-carrying wire falls off as $1/r$ in normal electromagnetism, here we expect the field of such a wire to asymptote to a constant (or perhaps grow logarithmically).  Thus, even though the particles in this theory are free to move along their appropriate one-dimensional subspace, it will be much harder to set these particles into motion than conventional charges, leading to large inductance associated with currents.

\subsection{Summary of Maxwell Equations}

The generalized Maxwell equations for the vector charge theory take the form:
\begin{equation}
\boxed{
\begin{split}
&\partial_i E^{ij} = \rho^j\\
&\partial_i B^{ij} = \tilde{\rho}^j \\
\epsilon_{iab}\epsilon_{jcd}\partial^a&\partial^c E^{bd} = \partial_t B_{ij} + \tilde{J}_{ij}\\
\epsilon_{iab}\epsilon_{jcd}\partial^a&\partial^c B^{bd} = -\partial_t E_{ij} - J_{ij}
\end{split}}
\end{equation}
Once again, note the nice symmetry between electric and magnetic quantities, reflecting the self-duality of the theory.

\section{Traceless Vector Charge Theory}

\subsection{Electrostatic Field}

Let us now move to the last of the rank 2 theories, which has the same Gauss's law as the previous case, $\partial_i E^{ij} = \rho^j$, but now with an extra trace condition, $E^i_{\,\,\,i} = 0$.  We start with a point source, $\partial_i E^{ij} = \rho^j \delta^{(3)}(r)$.  The corresponding electric field must once again take the form of Equation \ref{1dgeneral}.  The Gauss's law will lead to the same constraints as before, simplifying our electric field down to:
\begin{equation}
E^{ij} = \alpha\frac{(p^ir^j+r^ip^j)}{r^3} - \alpha \frac{(p\cdot r)\delta^{ij}}{r^3} + \bigg(\frac{3}{4\pi}-3\alpha\bigg)\frac{(p\cdot r)r^ir^j}{r^5}
\end{equation}
We must also impose the tracelessness condition, $E^i_{\,\,\,i} = (-4\alpha + \frac{3}{4\pi})(p\cdot r)/r^3 = 0$, which tells us that $\alpha = 3/16\pi$, so the electric field of a point charge has the form:
\begin{equation}
E^{ij} = \frac{3}{16\pi}\bigg(\frac{(p^ir^j+r^ip^j)}{r^3} - \frac{(p\cdot r)\delta^{ij}}{r^3} + \frac{(p\cdot r)r^ir^j}{r^5}\bigg)
\label{tracvecfield}
\end{equation}
Even before proceeding to the magnetostatic constraint, the electrostatic field of a point charge is already uniquely constrained in this theory (though one can check that the magnetostatic constraint is obeyed as well).

\subsection{Potential Formulation}

As always, we now seek some potential formulation for our electrostatic field.  Just as in the scalar charge traceless case, we shall not derive the potential directly from the magnetostatic condition, but rather will make an ansatz for the potential and then verify that it is the correct one.  We start with the potential formulation for the previous case, $E_{ij} = -\frac{1}{2}(\partial_i\phi_j + \partial_j\phi_i)$ and add in an appropriate term to make the electric field traceless:
\begin{equation}
E_{ij} = -\frac{1}{2}(\partial_i\phi_j + \partial_j\phi_i) + \frac{1}{3}\delta_{ij}(\partial_k\phi^k)
\end{equation}
Assuming the same general form for the potential as in Equation \ref{vecpot}, we find that Gauss's law is only satisfied if $\gamma' = 0$, $\alpha' = \frac{1}{16\pi}$, and $\beta = \frac{7}{16\pi}$, giving a potential:
\begin{equation}
\phi^j = \frac{1}{16\pi}\bigg(\frac{(p\cdot r)r^j}{r^3} + 7 \frac{p^j}{r}\bigg)
\end{equation}
The resulting electric field exactly matches what we found in Equation \ref{tracvecfield}.  Since the potential formulation works for the point charge, by linearity it will work for an arbitrary charge distribution.  Thus, this is the correct potential formulation, even though it has not been derived directly from the magnetostatic constraint.

The energy stored in the electric field of a static charge configuration is given by:
\begin{equation}
\begin{split}
&\epsilon = \frac{1}{2}\int E^{ij}E_{ij} = \\
\frac{1}{2}\int & E^{ij}(-\frac{1}{2}(\partial_i\phi_j + \partial_j\phi_i) + \frac{1}{3}\delta_{ij}(\partial_k\phi^k)) = \frac{1}{2}\int\rho^j\phi_j
\end{split}
\end{equation}
where we have integrated by parts and made use of Gauss's law and tracelessness.  Just as in the traceful theory, we find that $\phi_j$ can quite legitimately be regarded as the potential energy for charges $\rho^j$.

\subsection{Lorentz Force}

The fundamental vector charges in this theory are all fractonic and will have no sense of Lorentz forces.  Like the traceful vector charge theory, this theory will also have chiron bound states, carrying zero charge but nonzero charge angular moment $L^j$.  However, whereas the chirons in the traceful theory were fully mobile, in the present case the extra conservation laws restrict the chirons to be 1-dimensional particles, constrained to move only along the direction of their charge angular moment vector.  Projecting from the previous theory onto the appropriate one-dimensional subspace, the Lorentz force on a chiron becomes:
\begin{equation}
F^i = L^i\hat{L}^n\hat{L}^j\epsilon_{jk\ell}\partial^k E^{\ell n}
\end{equation}

\subsection{Currents and the Biot-Savart Law}

As in the traceless scalar charge theory, the microscopic current operator will take the form of a traceless symmetric tensor $J_{ij}$, representing the rate of hopping processes.  For this theory, the magnetic field tensor takes the following form:
\begin{equation}
B_{ij} = \epsilon_{i}^{\,\,\,k\ell}\partial_k(\tilde{B}_{j\ell}-\frac{1}{2}\delta_{j\ell}\tilde{B}^n_{\,\,\,n})
\end{equation}
where $\tilde{B}_{ij} = \epsilon_{i}^{\,\,\,ab}\epsilon_{j}^{\,\,\,cd}\partial_a\partial_cA_{bd}$ is the magnetic tensor from the traceful theory.  It is readily verified that this magnetic tensor is traceless, $B^i_{\,\,\,i} = 0$.  Also, this tensor does not look symmetric at first glance.  Nevertheless, we find:
\begin{equation}
\begin{split}
\epsilon^{ija}B_{ij} = (\delta^{kj}\delta^{a\ell} - \delta^{ka}\delta^{j\ell})\partial_k(\tilde{B}_{j\ell}-\frac{1}{2}\delta_{j\ell}\tilde{B}^n_{\,\,\,n}) =\\
\partial^j\tilde{B}_{j\ell} -\frac{1}{2}\partial_a\tilde{B}^i_{\,\,\,i} + \frac{1}{2}\partial_a\tilde{B}^i_{\,\,\,i} = 0
\end{split}
\end{equation}
where we have made use of $\partial^j\tilde{B}_{j\ell} = 0$.  This tells us that the antisymmetric component of $B_{ij}$ actually vanishes, so it is a symmetric traceless tensor, thereby allowing this theory to have a self-duality.  We can then write the magnetic tensor in manifestly symmetric form as:
\begin{equation}
B_{ij} = \frac{1}{2}(\epsilon_i^{\,\,\,k\ell}\partial_k\tilde{B}_{j\ell} + \epsilon_j^{\,\,\,k\ell}\partial_k\tilde{B}_{i\ell})
\end{equation}
In terms of the microscopic variable $A_{ij}$, we have the following unspeakable horror:
\begin{equation}
\begin{split}
B_{ij} = \frac{1}{2}\bigg(\epsilon_{jab}&(\partial_a\partial_k\partial_iA_{bk}-\partial_a\partial^2A_{bi}) + \\
&\epsilon_{iab}(\partial_a\partial_k\partial_jA_{bk}-\partial_a\partial^2A_{bj})\bigg)
\label{monster}
\end{split}
\end{equation}
The Hamiltonian has the standard form:
\begin{equation}
\int\bigg(\frac{1}{2}E^{ij}E_{ij} + \frac{1}{2}B^{ij}B_{ij} + J^{ij}A_{ij}\bigg)
\end{equation}
Plugging in the form for $B$ and making use of index symmetry, we obtain:
\begin{equation}
\begin{split}
\int\bigg(\frac{1}{2}E^{ij}E_{ij} + \frac{1}{2}\epsilon_{jab}(\partial_a\partial_k\partial_iA_{bk}-\partial_a\partial^2A_{bi})B^{ij} + \\
J^{ij}A_{ij}\bigg) = \\
\int\bigg(\frac{1}{2}E^{ij}E_{ij} - \frac{1}{2}A_{ij}\epsilon_{kai}\partial_a(\partial_j\partial_bB^{bk}-\partial^2B^{jk}) + \\
J^{ij}A_{ij}\bigg)
\end{split}
\end{equation}
Noting that $A_{ij}$ is symmetric, the time evolution equation for $E$ is given by:
\begin{equation}
\begin{split}
\frac{1}{2}\bigg(\epsilon_{kai}\partial^a(\partial_j\partial_bB^{bk}-\partial^2B_j^{\,\,\,k}) + &\epsilon_{kaj}\partial^a(\partial_i\partial_bB^{bk}-\partial^2B_i^{\,\,\,k})\bigg) \\
&= J_{ij} + \partial_t E_{ij}
\end{split}
\end{equation}
This Ampere's equation has an unusual feature, in that some of the terms on the left contain direct contractions between derivatives and the magnetic tensor, which can then be written in terms of the magnetic charge, $\partial_i B^{ij} = \tilde{\rho}^j$, as follows:
\begin{equation}
\begin{split}
\frac{1}{2}\bigg(\epsilon_{kai}\partial^a(\partial_j\tilde{\rho}^k-\partial^2B_j^{\,\,\,k}) + &\epsilon_{kaj}\partial^a(\partial_i\tilde{\rho}^k-\partial^2B_i^{\,\,\,k})\bigg) \\
&= J_{ij} + \partial_t E_{ij}
\end{split}
\end{equation}
It seems rather unusual at first to have magnetic charge appear in Ampere's equation, but there is nothing logically inconsistent about it.  (Actually, if we carefully examined Ampere's equation from the traceful vector charge theory, we would see magnetic charge appearing there as well.)  In particular, after taking a derivative, the magnetic charge terms are killed, and we still obtain the expected continuity equation for the electric charges:
\begin{equation}
\partial_t \rho^j +\partial_iJ^{ij} = 0
\end{equation}
In the absence of magnetic charge, and assuming a steady current configuration, we can drop some terms from Ampere's equation, which then simplifies to:
\begin{equation}
\partial_a\partial^2\bigg(\epsilon^{kai}B^j_{\,\,\,k} + \epsilon^{kaj}B^i_{\,\,\,k}\bigg) = -2J^{ij}
\end{equation}
We can write a solution as:
\begin{equation}
\begin{split}
\epsilon^{kai}B^j_{\,\,\,k} &+ \epsilon^{kaj}B^i_{\,\,\,k} = \\
&\bigg(\frac{1}{4\pi}\int dr' J^{ij}(r')\frac{(r-r')^a}{|r-r'|}\bigg) + \epsilon^{abc}\partial_b\lambda_{c}^{\,\,\,ij}
\end{split}
\end{equation}
Inverting for $B^{ij}$ gives:
\begin{equation}
\begin{split}
B^{ij} =& \\
\bigg(\frac{1}{12\pi}&\int dr' J^{ik}(r')\epsilon^{j\ell k}\frac{(r-r')^\ell}{|r-r'|}\bigg) + \frac{1}{3}(\partial_\ell\lambda^{i\ell j} - \partial^i\lambda_\ell^{\,\,\,\ell j})
\end{split}
\end{equation}
The first term by itself is not symmetric.  However, we can choose:
\begin{equation}
\lambda^{i\ell j} = \frac{1}{4\pi}\int dr'J^{jk}(r')\epsilon^{i\ell k} |r-r'|
\end{equation}
The resulting magnetic field tensor is:
\begin{equation}
B^{ij} = \frac{1}{12\pi}\int dr'(J^{ik}(r')\epsilon^{j\ell k} + J^{jk}(r')\epsilon^{i\ell k})\frac{(r-r')^\ell}{|r-r'|}
\end{equation}
This equation is manifestly traceless and symmetric.  It also satisfies $\partial_iB^{ij} = 0$ (making use of $\partial_iJ^{ij} = 0$ for a steady current), which indicates the absence of magnetic charge.  We have therefore found the correct Biot-Savart law for this theory, which can then be used to obtain the magnetic field tensor for an arbitrary steady current configuration.  Note that, like in the previous theory, the Biot-Savart law falls off very slowly, leading to large inductances in this theory.

\subsection{Summary of Maxwell Equations}

The generalized Maxwell equations for the traceless vector charge theory take the following mildly nauseating form:
\begin{equation}
\boxed{
\begin{split}
&\partial_iE^{ij} = \rho^j \\
&\partial_iB^{ij} = \tilde{\rho}^j \\
\frac{1}{2}\bigg(\epsilon_{iak}\partial^a(\partial_j&\tilde{\rho}^k-\partial^2B_j^{\,\,\,k}) + \epsilon_{jak}\partial^a(\partial_i\tilde{\rho}^k-\partial^2B_i^{\,\,\,k})\bigg) \\
&=  - \partial_t E_{ij} - J_{ij} \\
\frac{1}{2}\bigg(\epsilon_{iak}\partial^a(\partial_j&\rho^k-\partial^2E_j^{\,\,\,k}) + \epsilon_{jak}\partial^a(\partial_i\rho^k-\partial^2E_i^{\,\,\,k})\bigg) \\
&= \partial_t B_{ij} + \tilde{J}_{ij} \\
&E^i_{\,\,\,i} = 0\\
&B^i_{\,\,\,i} = 0
\end{split}}
\end{equation}
Once again, the equations have a nice electric-magnetic symmetry, reflecting the self-duality of the theory.

\section{Conclusion}

In this work, we have generalized some of the basic notions of electromagnetism to systems with tensor $U(1)$ gauge fields, instead of the conventional $U(1)$ vector gauge theory.  The topics treated here have included electrostatic fields, potential formulations, Maxwell equations, Lorentz forces, and Biot-Savart laws.  There is much that carries over quite naturally, while some concepts have interesting modifications.  While we have laid the groundwork here, there is plenty more that could be done.  Obviously one could work out the electromagnetic properties of rank 3 and higher theories.  But also, electromagnetism is a much broader subject than just the topics treated here.  In principle, one could pull out their favorite electromagnetism textbook and generalize everything chapter by chapter to the higher rank analogue.  One could work out the theory of higher rank radiation, higher rank circuits, higher rank waveguides, and so on.  It's a brand new playground.

\section*{Acknowledgments}

I would like to particularly thank Sagar Vijay, for numerous helpful discussions and for pointing out an error regarding one of the magnetic tensors used in previous work.  I would also like to thank Senthil Todadri, Liujun Zou, Yahui Zhang, and Inti Sodemann for useful discussions.  This work was supported by NSF DMR-1305741.

\section*{Appendix A: Magnetic Particles in the Scalar Charge Theory}

As mentioned in the text, the scalar charge theory (without trace condition) is not self-dual.  Therefore, we cannot simply read off the fields for magnetic particles from the corresponding electric ones, and they must be calculated separately.  While the electric charges of this theory were scalars, the magnetic charges are vectors, $\partial_iB^{ij} = \rho^j$.  For an electrostatic configuration, in the absence of electric currents, we must also have:
\begin{equation}
\epsilon^{iab}\partial_a B_b^{\,\,\,j} + \epsilon^{jab}\partial_a B_b^{\,\,\,i} = 0
\end{equation}
Recall that $B^{ij}$ in this theory is traceless, but not symmetric.  For brevity, we will go right for the throat and take a potential formulation:
\begin{equation}
B_{ij} = \partial_i\tilde{\phi}_j - \frac{1}{3}\delta_{ij}(\partial_k\tilde{\phi}^k)
\end{equation}
which can be verified to solve the electrostatic constraint.  We now force this magnetic field to satisfy the magnetic Gauss's law for a point source:
\begin{equation}
\partial_iB^{ij} = \partial^2 \tilde{\phi}^j - \frac{1}{3}\partial^j(\partial_k\tilde{\phi}^k) = p^j\delta^{(3)}(r)
\end{equation}
As an educated guess, we will take an ansatz of the form:
\begin{equation}
\tilde{\phi}^j = \alpha \frac{p^j}{r} + \beta\frac{(p\cdot r)r^j}{r^3}
\end{equation}
We have:
\begin{equation}
\partial^2 \tilde{\phi}^j = -\frac{4\pi}{3}(3\alpha + \beta) p^j\delta^{(3)}(r) + 2\beta\bigg(\frac{p^j}{r^3} - 3 \frac{(p\cdot r)r^j}{r^5}\bigg)
\end{equation}
\begin{equation}
\partial^j(\partial_k\tilde{\phi}^k) = (\beta-\alpha)\bigg(\frac{p^j}{r^3} - 3\frac{(p\cdot r)r^j}{r^5} +\frac{4\pi}{3}p^j\delta^{(3)}(r)\bigg)
\end{equation}
\begin{equation}
\begin{split}
\partial_iB^{ij} &= \\
-\frac{4\pi}{9}&(8\alpha + 4\beta)p^j\delta^{(3)}(r) + \frac{1}{3} (5\beta + \alpha)\bigg(\frac{p^j}{r^3}-3\frac{(p\cdot r)r^j}{r^5}\bigg)
\end{split}
\end{equation}
To solve the magnetic Gauss's law, we need $\alpha = -5\beta$ and $16\pi\beta = 1$, which has the solution $\alpha = -\frac{5}{16\pi}$ and $\beta = \frac{1}{16\pi}$.  Our potential then has the form:
\begin{equation}
\tilde{\phi}^j = -\frac{1}{16\pi}\bigg(5\frac{p^j}{r} - \frac{(p\cdot r)r^j}{r^3}\bigg)
\end{equation}
and the magnetic field takes the form:
\begin{equation}
\begin{split}
B^{ij} =& \\
-\frac{1}{16\pi}&\bigg(-5\frac{p^jr^i}{r^3} - \frac{p^ir^j}{r^3} + \frac{(p\cdot r)\delta^{ij}}{r^3} + 3\frac{(p\cdot r)r^ir^j}{r^5}\bigg)
\end{split}
\end{equation}
We can also consider a traceless (non-symmetric) tensor current $\tilde{J}^{ij}$ for the magnetic particles.  In the presence of this magnetic current, the time evolution equation for $B_{ij}$ is modified to:
\begin{equation}
\epsilon^{iab}\partial_aE_b^{\,\,\,j} = \partial_tB^{ij} + \tilde{J}^{ij}
\end{equation}
Like in the electric case, this equation can be used as the fundamental definition of the current tensor $\tilde{J}^{ij}$.  Taking a derivative yields the continuity equation:
\begin{equation}
\partial_t \tilde{\rho}^j + \partial_i\tilde{J}^{ij} = 0
\end{equation}
A steady magnetic current will generate a time-independent electric field, which must satisfy:
\begin{equation}
\partial_a(\epsilon^{iab}E_b^{\,\,\,j}) = \tilde{J}^{ij}
\end{equation}
which we can solve via:
\begin{equation}
\epsilon^{iab}E_b^{\,\,\,j} = \bigg(\frac{1}{4\pi}\int dr' \tilde{J}^{ij}(r') \frac{(r-r')^a}{|r-r'|^3}\bigg) + \epsilon^{abc}\partial_b\lambda_{cji}
\end{equation}
\begin{equation}
E^{ij} = \frac{1}{8\pi}\int dr' \tilde{J}^{\,\,\,j}_{k}(r')\epsilon^{k\ell i}\frac{(r-r')_\ell}{|r-r'|^3} + \frac{1}{2}(\partial^j \lambda^{\ell\,\,\,\,\,i}_{\,\,\,\ell} - \partial^\ell\lambda^{j\,\,\,\,\,i}_{\,\,\,\ell})
\end{equation}
We then choose:
\begin{equation}
\lambda^{j\ell i} = -\frac{1}{4\pi}\int dr'\tilde{J}^{\,\,\,i}_{k}(r')\epsilon^{k\ell j}\frac{1}{|r-r'|}
\end{equation}
The resulting electric field is:
\begin{equation}
E^{ij} = \frac{1}{8\pi}\int dr' (\tilde{J}_k^{\,\,\,j}(r')\epsilon^{k\ell i} + \tilde{J}_k^{\,\,\,i}(r')\epsilon^{k\ell j})\frac{(r-r')_\ell}{|r-r'|^3}
\end{equation}
This electric field is symmetric and also satisfies $\partial_i\partial_j E^{ij} = 0$, indicating the absence of electric charge.  We have therefore found the correct dual Biot-Savart law for this theory.  From this formula, we can construct the electric field tensor corresponding to an arbitrary steady current of magnetic particles.

In order to derive the Lorentz force for the magnetic particles, it is useful to formulate in terms of the dual gauge variable $\tilde{A}_{ij}$, which is canonically conjugate to $B_{ij}$.  Note that $\tilde{A}_{ij}$ is non-symmetric.  From the Ampere's equation of this theory, in the absence of electric currents, we can reverse engineer the following expression for $E^{ij}$ in terms of the dual potential:
\begin{equation}
E^{ij} = -\frac{1}{2}(\epsilon^{iab}\partial_a\tilde{A}_b^{\,\,\,j} + \epsilon^{jab}\partial_a\tilde{A}_b^{\,\,\,i})
\end{equation}
We note that $\tilde{A}_{ij}$ represents the phase associated with a $j$ oriented charge hopping in the $i$ direction.  Therefore the effective dual vector potential for a charge $p^j$ is given by $\tilde{A}_i^{eff} = p^j\tilde{A}_{ij}$.  The role of the ``magnetic field" for this particle (in the sense of the conventional Lorentz force) will be played by $p_i\hat{p}_j(\hat{p}^k\epsilon^{jnm}\partial_n\tilde{A}_{mk}) = -p_i(\hat{p}_j\hat{p}_kE^{jk})$.  The Lorentz force then takes the form:
\begin{equation}
F^i = p_j(P^{ik}B_{kj} - \epsilon^{ik\ell}v_k\hat{p}_\ell\hat{p}_mE^{jm})
\end{equation}
where $P^{ik}$ is the projector into the plane transverse to $p^j$.  We note that this force always lies in the transverse plane, consistent with the 2-dimensional nature of the particles.

\section*{Appendix B: Useful Formulas}

As we all learned in our undergraduate days, the divergence of the electric field of a point charge is given by:
\begin{equation}
\partial_i\frac{r^i}{r^3} = 4\pi\delta^{(3)}(r)
\end{equation}
This can be derived by performing an integral over a ball $S$ of radius $R$ centered at the origin:
\begin{equation}
\int_S \partial_i \frac{r^i}{r^3} = \int_{\partial S}\frac{1}{R^2} = 4\pi
\end{equation}
Since this must be true for arbitrary $R$, we can conclude that there must be a delta function contribution at the origin.  We will list below some further useful formulas, derived via the same technique, which are useful for the manipulations performed in the main text.
\begin{equation}
\partial^i\frac{r^j}{r^3} = \frac{4\pi}{3}\delta^{ij}\delta^{(3)}(r) + \frac{\delta^{ij}}{r^3} - 3\frac{r^ir^j}{r^5}
\end{equation}
\begin{equation}
\partial_k\frac{r^ir^jr^k}{r^5} = \frac{4\pi}{3}\delta^{ij}\delta^{(3)}(r)
\end{equation}
\begin{equation}
\begin{split}
\partial_\ell \frac{r^ir^jr^k}{r^5} = \frac{4\pi}{15}(\delta^{ij}\delta^{k\ell}+\delta^{ik}\delta^{j\ell}+\delta^{i\ell}\delta^{jk}) \delta^{(3)}(r) +\\
 \frac{\delta^{\ell i}r^jr^k + \delta^{\ell j}r^ir^k + \delta^{\ell k}r^ir^j}{r^5} - 5\frac{r^ir^jr^kr^\ell}{r^7}
\end{split}
\end{equation}
\begin{equation}
\partial^k \frac{r^ir^j}{r^3} = \frac{\delta^{ik}r^j + \delta^{jk}r^i}{r^3} - 3\frac{r^ir^jr^k}{r^5}
\end{equation}
\begin{equation}
\partial^2 \frac{r^ir^j}{r^3} = -\frac{4\pi}{3}\delta^{ij}\delta^{(3)}(r) +2\bigg(\frac{\delta^{ij}}{r^3} - 3\frac{r^ir^j}{r^5}\bigg)
\end{equation}
\begin{equation}
\partial^i\partial_k \frac{r^kr^j}{r^3} = \frac{4\pi}{3}\delta^{ij}\delta^{(3)}(r) + \frac{\delta^{ij}}{r^3} - 3\frac{r^ir^j}{r^5} = \partial^i\frac{r^j}{r^3}
\end{equation}

\section*{Appendix C: Generalized Curl Constraints}

In the text, we have considered different forms of Gauss's laws.  These were all either generalized divergences or trace conditions.  One might also consider generalized curl constraints, such as $\epsilon^{ijk}\partial_jE_k^{\,\,\,\ell} = \rho^{i\ell}$?  To see why we have not analyzed this type of theory, we first examine the rank 1 analogue, $\vec{\nabla} \times \vec{E} = \vec{\rho}$.  Whereas the divergence constraint gave rise to point particles, such a curl constraint naturally gives rise to string-like excitations, due to the fact that:
\begin{equation}
\partial_i \rho^i = \epsilon^{ijk}\partial_i\partial_jE_k = 0
\end{equation}
This constraint on the vector charges automatically forces them to form closed loops, a constraint which cannot be broken within the Hilbert space.  Another way to understand this is to note that we can rewrite the gauge constraint as:
\begin{equation}
\rho^i = \partial_j\epsilon^{ijk}E_k = \partial_j \tilde{E}^{ij}
\end{equation}
where we have defined the antisymmetric tensor $\tilde{E}^{ij} = \epsilon^{ijk}E_k$, which captures all the information of the original vector.  We therefore see that a curl constraint on $E_i$ actually gives us a dual formulation of an antisymmetric $U(1)$ tensor gauge field (the standard Kalb-Ramond theory).  The generalized electromagnetism of this phase is therefore inherently more complicated, since one must think in terms of a closed string theory instead of a theory of point particles.  For the purposes of describing three-dimensional spin liquids, the analysis is a moot point anyway, as Kalb-Ramond theory is unstable to confinement in three spatial dimensions due to instanton effects\cite{analytic,numerical,gerbe}.  This theory therefore does not exist as a stable phase of matter.

Similarly, if we take a ``curl" gauge constraint on our tensor gauge field, say $\epsilon^{ijk}\partial_j E_k^{\,\,\,\ell} = \rho^{i\ell}$, the charges would obey the constraint:
\begin{equation}
\partial_i\rho^{i\ell} = 0
\end{equation}
This forces the tensor charges to line up along one-dimensional string-like structures.  Similar stories hold for other curl constraints.  Therefore, the charges of all of these theories are extended objects, not point particles, which makes the analysis of the generalized electromagnetism much more complicated.  Furthermore, such putative phases may be destabilized by instantons, so it is not clear whether these correspond to stable phases of matter at all.  We therefore defer a more detailed analysis of such theories to future work.

\section*{Appendix D: Microscopic Models}

In the main text, we have mostly abstracted from the microscopic behavior of the higher rank $U(1)$ spin liquids, instead relying on a more macroscopic field-theoretic approach.  This is a useful point of view, since most of the essential physics is independent of the microscopics, with a few small exceptions (such as the precise quantization of dipole moments in the scalar charge theory).  Nevertheless, it is useful to keep the microscopic theories in mind, since they often offer important clues in elucidating the physical principles of generalized electromagnetism.  We will therefore review here some of the basic principles of the previously discovered lattice models which are known to exhibit the behavior of higher rank $U(1)$ spin liquids.\cite{alex,cenke1,cenke2}

\begin{figure}[t!]
 \centering
 \includegraphics[scale=0.25]{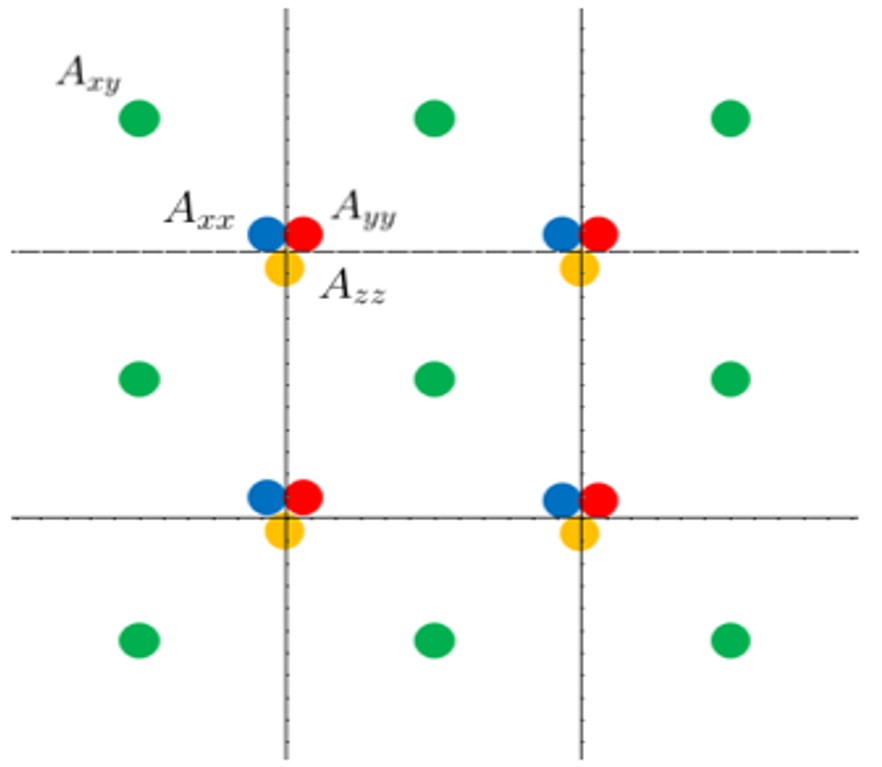}
 \caption{The microscopic model takes the form of a lattice rotor model, where each independent component of the tensor corresponds to a separate rotor.  Diagonal components of the tensor live on each site of a cubic lattice (a 2d cross-section of which is pictured above).  Off-diagonal components live on the appropriate plaquettes of the lattice, with $A_{xy}$ on plaquettes in the $xy$ plane, for example.}
 \label{fig:lattice}
 \end{figure}

For normal vector gauge theories, constructing lattice models is a simple matter of letting the gauge field live on links of the lattice, with $A_x$ living on $x$-directed links, $A_y$ living on $y$-directed links, and so on.  It is less obvious how one should put the six components of a rank 2 tensor $A_{ij}$ on the lattice.  The key piece of intuition is to look at the simplest rank 2 tensor: a second derivative $\partial_i\partial_j\alpha$ of a scalar $\alpha$.  If we choose $\alpha$ to live on the sites of the lattice, then diagonals like $\partial_x\partial_x\alpha$ will also live on sites, while off-diagonals like $\partial_x\partial_y\alpha$ will live on plaquettes in the appropriate plane.  We therefore can construct simple cubic lattice models for rank 2 theories by allowing $A_{xx}$, $A_{yy}$, and $A_{zz}$ to live on the sites of the lattice, while $A_{xy}$, $A_{xz}$ and $A_{yz}$ live on the appropriate plaquettes ($A_{xy}$ in the $xy$-plane, and so on).  This is illustrated in Figure \ref{fig:lattice}.  Similar stories will hold for theories of even higher rank.  We also generically allow the gauge field to be compact, identifying $A_{ij}\sim A_{ij} + 2\pi$, making each component of the tensor a quantum rotor.  The corresponding electric tensor components $E_{ij}$ then become the angular momenta of these rotors.

All of the $U(1)$ theories of a given rank can be constructed from the same basic lattice degrees of freedom.  They are distinguished, however, by their gauge constraint structure, as determined by the Hamiltonian of the system.  For example, let us work through a lattice model of the scalar charge theory.  Microscopic models for the other higher rank theories are a straightforward generalization.

The most important term in the Hamiltonian of the system is a ``$U$" term, which energetically imposes the gauge constraint.  For the scalar charge theory, this takes the form:
\begin{equation}
H_U = U(\partial_i\partial_jE^{ij})^2
\end{equation}
where the indices run over $\{x,y,z\}$ and $U$ is a large positive number.  This corresponds to a generalized type of ``spin-ice rule," constraining how the rotor momenta line up relative to their neighbors.  In the low-energy sector, states will obey $\partial_i\partial_jE^{ij} = 0$, which then implies gauge invariance under the following transformations:
\begin{equation}
A_{ij}\rightarrow A_{ij} + \partial_i\partial_j\alpha
\end{equation}
for gauge parameter $\alpha$ with arbitrary spatial dependence.

Our low-energy Hamiltonian should also feature the most relevant terms which are consistent with this gauge transformation ($i.e.$ commute with the gauge constraint).  It is straightforward to check that the resulting Hamiltonian takes the form:
\begin{equation}
H = \frac{1}{2}(gE^{ij}E_{ij} + B^{ij}B_{ij}) + U(\partial_i\partial_jE^{ij})^2
\end{equation}
where $g$ is a numerical coefficient.  (The coefficient of the $B$ term is normalized to 1 for convenience.)  The magnetic tensor $B_{ij}$ takes the form $B_{ij} = \epsilon_{iab}\partial^a A^b_{\,\,\,j}$.  All other possible terms in the Hamiltonian have larger numbers of derivatives and are irrelevant to the low-energy physics.

At slightly higher energies, there are also states in the Hilbert space which do not obey the gauge constraint, $\partial_i\partial_jE^{ij}\neq 0$.  In this case, we define a charge density as:
\begin{equation}
\rho = \partial_i\partial_jE^{ij}
\end{equation}
which lives on the sites of the lattice.  These charges obey both conservation of charge:
\begin{equation}
\sum_{sites}\rho = \textrm{constant}
\end{equation}
and conservation of dipole moment:
\begin{equation}
\sum_{sites}\rho \vec{x} = \textrm{constant}
\end{equation}
These lattice conservation laws imply that the charges in this lattice model are fractons, as discussed in the main text.

Note that an individual rotor operator $e^{iA_{ij}}$, which raises $E_{ij}$ of the rotor by 1, does not commute with the gauge constraint and will therefore create and/or move particles.  Any such operator must respect both conservation of charge and dipole moment.  It can therefore only create charges in quadrupolar configurations, or equivalently, jointly move dipolar bound states.  It can easily be checked that $e^{iA_{ij}}$ will move an $i$ directed dipole in the $j$ direction, as discussed in the main text.

The basic principles discussed here easily transfer over to writing down microscopic models for any of the higher rank $U(1)$ spin liquids.  One first writes down the appropriate ``$U$" term enforcing a gauge constraint.  Then one writes down the most relevant terms which commute with this gauge constraint.  In general, the Hamiltonian will schematically have an ``$E^2+B^2$" form and will correspond to a stable phase of matter in $(3+1)$ dimensions.\cite{alex}

\end{document}